\documentclass[psfig]{mn2e}

\usepackage{times}

\newcommand{\bgr}{\bibitem[\protect\citename{dummy }1893]{dum}}

\begin{document}
\title[Galaxy formation around high-redshift QSOs]
{An excess of star-forming galaxies in the fields of high-redshift QSOs}{}

\author[J. A. Stevens et al.]
{J.\ A.\ Stevens,$^{1}$ Matt\ J.\ Jarvis,$^{1}$ K.\ E.\ K.\ Coppin,$^2$ M.\ J.\ Page,$^3$
  T.\ R.\ Greve,$^4$ \newauthor F.\ J.\ Carrera$^{5}$ and R.\ J.\ Ivison$^{6,7}$
\\
$^1$ Centre for Astrophysics Research, University of Hertfordshire, College
Lane, Herts AL10 9AB\\
$^2$ Institute for Computational Cosmology, Durham University, South Road,
Durham DH1 3LE \\ 
$^3$ Mullard Space Science Laboratory, University College London, Holmbury
St. Mary, Dorking, Surrey, RH5 6NT  \\
$^4$ Max-Planck Institute f\"{u}r Astronomie, K\"{o}nigstuhl 17, Heidelberg
69117, Germany\\
$^5$ Instituto de Fisica de Cantabria (CSIC-UC), Avenida de los Castros
39005 Santander, Spain\\
$^6$ Astronomy Technology Centre, Royal Observatory, Blackford Hill,
Edinburgh EH9 3HJ \\
$^7$ Scottish Universities Physics Alliance, Institute for Astronomy,
University of Edinburgh, Royal Observatory, Blackford Hill, Edinburgh EH9 3HJ
}
\date{draft 3.0}

\maketitle \begin{abstract} 
\noindent We present submillimetre and mid-infrared imaging observations of five fields centred on quasi-stellar objects (QSOs) at $1.7<z<2.8$. All $5$ QSOs were detected previously at submillimetre wavelengths. At 850 (450)~$\mu$m we detect 17 (11) submillimetre galaxies (SMGs) in addition to the QSOs.  The total area mapped at 850~$\mu$m is $\sim28$~arcmin$^2$ down to RMS noise levels of $1-2$ mJy/beam, depending on the field.  Integral number counts are computed from the 850~$\mu$m data using the same analytical techniques adopted by `blank-field' submillimetre surveys.  We find that the `QSO-field' counts show a clear excess over the blank-field counts at deboosted flux densities of $\sim2-4$ mJy; at higher flux densities the counts are consistent with the blank-field counts. Robust mid-infrared counterparts are identified for all four submillimetre detected QSOs and $\sim60$ per cent of the SMGs. The mid-infrared colours of the QSOs are similar to those of the local ULIRG/AGN Mrk~231 if placed at $1<z<3$ whilst most of the SMGs have colours very similar to those of the local ULIRG Arp~220 at $1<z<3$. Mid-infrared diagnostics therefore find no strong evidence that the SMGs host buried AGN although we cannot rule out such a possibility. Taken together our results suggest that the QSOs sit in regions of the early universe which are undergoing an enhanced level of major star-formation activity, and should evolve to become similarly dense regions containing massive galaxies at the present epoch. Finally, we find evidence that the level of star-formation activity in individual galaxies appears to be lower around the QSOs than it is around more powerful radio-loud AGN at higher redshifts.
\end{abstract}

%
\footnotetext{$^*$E-mail: j.a.stevens@herts.ac.uk}

\section{Introduction}

The popular hierarchical model of galaxy formation predicts that elliptical galaxies found today in the cores of rich clusters formed at high redshifts and at rare high-density peaks of the dark matter distribution. Within these regions, gas-rich proto-galaxies merge together rapidly, forming stars at a high rate. The same reservoir of gas used to build the stellar mass can also fuel the growth of the supermassive black holes (SMBH) found dormant in the centre of the galaxies at low redshift (e.g. Kauffmann \& Haehnelt 2000; Page et al. 2001; Hopkins et al. 2008). Since massive star formation is known to be a dusty phenomenon, the optical light from many young galaxies should be highly obscured. Dust enshrouded star formation is, however, a luminous phenomenon in the millimetre through far-infrared wavebands where the star-light absorbed by dust grains is re-emitted. Indeed, in the last decade, deep surveys at submillimetre (e.g. Smail, Ivison \& Blain 1997; Hughes et al. 1998; Eales et al. 1999; Scott et al. 2002; Coppin et al. 2006) and millimetre (e.g. Greve et al. 2004; Bertoldi et al. 2007) wavelengths have identified a high-redshift population of massive, dusty galaxies that are undergoing extreme starbursts (e.g. Smail et al. 2002).  Likewise, hotter dust heated in the circumnuclear environment of an AGN will emit at mid-infrared wavelengths. Therefore, submillimetre and mid-infrared observations can be combined to study activity in forming galaxies due to dust obscured starbursts and AGN.

To-date there has been no published survey at (sub)millimetre or far-infrared wavelengths of a randomly selected region of sky over an area sufficiently large to ensure that very rare structures in the early universe are contained within it. Although large surveys of this type will be available soon from instruments such as SCUBA-2 on the JCMT and SPIRE/PACS on-board the {\em Herschel Space Observatory\/}, published surveys have so far been limited to an area of $\sim0.5$~deg$^{-2}$ (Coppin et al. 2006; Austermann et al. 2010; Wei\ss{} et al. 2009). One of the largest published submillimetre surveys is the SCUBA HAlf-Degree Extragalactic Survey (SHADES; Mortier et al. 2005; Coppin et al. 2006) which, due to technical problems, mapped only 0.2 deg$^2$ discovering $\sim100$ submillimetre galaxies (SMGs). Whilst this survey provided one of the largest and most uniformly selected samples of SMGs ever assembled, the area of sky surveyed is extremely small by most standards. For example, $\Lambda$CDM simulations predict that a dark matter halo of mass greater than a few times $10^{14}$~M$_{\odot}$ at $z=2$ will evolve to become a very rich cluster of galaxies by $z=0$ (e.g. Farrah et al. 2006). However, since the surface density of such massive haloes at $z\geq2$ is only of order $0.01$~deg$^{-2}$ (Evrard et al. 2002), a wide-area survey would be required in order to contain one by chance.

One promising method of locating rare structures is to target the fields of high-redshift active galactic nuclei (AGN). Given their huge luminosities, such objects must contain a SMBH even at very high redshifts, and therefore must represent some of the most massive objects in existence at their epochs. They should thus act as signposts to rare high-density regions of the early universe, and can be observed efficiently with existing technology. Several groups have exploited this technique to search for proto-clusters forming around high-redshift radio galaxies (e.g. Kurk et al. 2000; Pentericci et al. 2000; Ivison et al. 2000; Stevens et al. 2003; Smail et al. 2003; De Breuck et al. 2004; Greve et al. 2007; Venemans et al. 2007) and quasi-stellar objects (QSOs; Stevens et al. 2004; Priddey, Ivison \& Isaak 2008). Most of these studies have concentrated on extreme objects, those with the highest luminosities and at the highest redshifts, and all have found over-densities of star-forming galaxies (typically Ly$\alpha$ emitters or SMGs) in the fields of the AGN. Here we extend this work to investigate the environments of typical QSOs at submillimetre and mid-infrared wavelengths. The QSOs are selected over the redshift range $1.7<z<2.8$ where the star-formation rate density and accretion luminosity density of the universe peak.  This work is motivated by the desire to study the coupled formation of stellar mass and black holes in galaxies that should evolve to become typical ellipticals in moderately rich clusters by the present day. Since the bulk of the field SMG population are also found at these redshifts (Chapman et al. 2003, 2005) we can also compare the properties of galaxies found in different environments.

This paper presents submillimetre and mid-infrared imaging data for 5 QSO fields. We extract submillimetre catalogues and integral number counts, and compare these with results from blank field surveys. We then characterise the mid-infrared properties of the QSOs and the SMGs discovered in their fields.

A Hubble constant $H_0=70$\ km\,s$^{-1}$\,Mpc$^{-1}$ and density parameters $\Omega_{\Lambda}=0.7$ and $\Omega_{\rm m}=0.3$ are assumed throughout.

\section{Sample selection}

\begin{table*}
\footnotesize
\centering
\def\baselinestretch{1}                                     
\caption[dum]{\small The sample of QSOs mapped with
  SCUBA. Coordinates are those of the optical QSO. Absorption corrected X-ray
  luminosities are given for the $0.5-2$~keV band whilst far-infrared
  luminosities are calculated from submillimetre photometric observations at
  850~$\mu$m (see Stevens et al. 2005 for more details on how these
  luminosities were calculated).} 
\label{table:sample}
\vspace*{0.1in}
\begin{tabular}{lccccc} \hline
Name & RA & Dec & $z$ & log($L_x$) & log(L$_{\rm FIR}$) \\
& (J2000.0) & (J2000.0) & &[log (erg\ s$^{-1}$)] &  [log (L$_{\odot}$)] \\ \hline
RX\,J$005734.78-272827.4$ & $00^{\rm h}\ 57^{\rm m}\ 34^{\rm s}.94$ & $-27^{\circ}\
       28^{\prime}\ 28^{\prime\prime}.0$ & $2.19$ & $44.85^{+0.14}_{-0.14}$ &
       $13.32^{+0.08}_{-0.06}$ \\ [5pt]
RX\,J$094144.51+385434.8$ & $09^{\rm h}\ 41^{\rm m}\ 44^{\rm s}.61$ & $+38^{\circ}\
       54^{\prime}\ 39^{\prime\prime}.1$ & $1.82$ & $44.83^{+0.12}_{-0.14}$ &
       $13.41^{+0.05}_{-0.07}$ \\ [5pt] 
RX\,J$121803.82+470854.6$ & $12^{\rm h}\ 18^{\rm m}\ 04^{\rm s}.54$ & $+47^{\circ}\
       08^{\prime}\ 51^{\prime\prime}.0$ & $1.74$ & $44.69^{+0.17}_{-0.15}$ &
       $13.10^{+0.08}_{-0.10}$\\ [5pt] 
RX\,J$124913.86-055906.2$ & $12^{\rm h}\ 49^{\rm m}\ 13^{\rm s}.85$ & $-05^{\circ}\
       59^{\prime}\ 19^{\prime\prime}.4$ & $2.21$ & $45.10^{+0.10}_{-0.13}$ &
       $13.11^{+0.09}_{-0.11}$\\ [5pt]
RX\,J$163308.57+570258.7$ & $16^{\rm h}\ 33^{\rm m}\ 08^{\rm s}.59$ & $+57^{\circ}\
       02^{\prime}\ 54^{\prime\prime}.8$ & $2.80$ & $45.24^{+0.11}_{-0.10}$ &
       $13.00^{+0.08}_{-0.11}$\\
\hline
\end{tabular}
\end{table*}

Our targets are drawn from the sample of 28 hard X-ray selected QSOs described by Page, Mittaz \& Carrera (2001). The objects have X-ray spectra harder than the X-ray background, consistent with moderate absorption ($20.5<\rm{log}(N_{\rm H}/\rm{cm}^{-2})<23$) and display broad optical emission lines.  In previous work (Page et al. 2001; Stevens et al. 2005) we have observed $20$ of these objects using SCUBA in photometry mode. The remaining $8$ objects were not targetted with SCUBA due to observing constraints so this sample should be representative of the population.  Our sample for subsequent mapping comprises those QSOs with photometry mode detections of $>5$~mJy/beam. This flux density threshold is based on limitations imposed by current sensitivity limits and by confusion noise, but also means that any source detected at $z>1$ must be either an ultraluminous or hyperluminous far-infrared galaxy if the emission is from dust, and if the temperature of this dust is typical of such objects detected at low redshift.  The observations of RX~J$094144.51+385434.8$ were published by Stevens et al. (2004) but are included here for completeness. The 5 QSOs are listed in Table~\ref{table:sample} along with some basic parameters.  They are selected to (1) span the redshift range $1.7<z<2.8$ and (2) have $0.5-2$\ keV X-ray luminosities within $\sim0.7$ dex of $L_X^*$, the break in the X-ray luminosity function.

\section{Observations and data reduction}

\subsection{SCUBA observations}

SCUBA jiggle-map observations at 450 and 850~$\mu$m were made at the JCMT
between 2003 July and 2004 December. These observations are summarized in
Table~\ref{table:log}. We used a 64-point jiggle pattern to fully sample the
image plane at both wavelengths but employed a number of different chop/nod
strategies.  For RX~J$163308$, RX~J$094144$ and RX~J$121803$ we chopped and
nodded East-West with throws of $30$, $30$ and $45$~arcsecs respectively. For
these objects the off-positions fall at a fixed position on the sky; a real
source thus has negative images of itself to the East and West, separated by
twice the chop throw but with half the amplitude.  For RX~J$005734$ and
RX~J$124913$ we chopped and nodded in azimuth by 30 arcsec so the off-positions
rotate with the field throughout the observation. 

Each mapping observation, taking typically one hour to complete, was bracketed
by pointing observations on bright point-like objects.  Pointing offsets were
in most cases $<2$~arcsec and in all cases $<3$~arcsec. Focus was checked
throughout the night but particularly around sunset and sunrise when rapid
changes are known to occur. 

Atmospheric transmission was determined with skydips and checked against the
JCMT water vapour radiometer. Maps of Uranus, Mars and/or CRL618 were used for
flux density calibration while maps of bright blazars were used to measure the
beam profile. At 850-$\mu$m the calibration was found to be stable enough to
apply a single calibration to all observations; our combined flux calibration
factor (FCF) was $221\pm15$ Jy/V/beam giving a calibration uncertainty of about
$7$ per cent. At 450-$\mu$m the FCF varies with the atmospheric conditions
because the filter profile contains a relatively strong water line, but also
depends quite strongly on the thermal state of the antenna (see Jenness et
al. 2002). At 450~$\mu$m we thus applied our best estimate of the FCF to each
observation in turn; FCFs varied from $220$ to $490$ Jy/V/beam. The calibration
uncertainty at 450~$\mu$m is $\sim20$ per cent.

Data were reduced with standard {\sc starlink} {\sc surf} routines up to the
rebinning stage (extinction correction, flat-fielding and despiking). For each
mapping observation the individual bolometer timestreams were inspected and
noisy integrations/noisy bolometers were blanked.  We then used adapted
versions of the {\sc surf} tasks {\sc setbolwt} and {\sc rebin} to make
bolometer noise weighted mean signal and noise maps on a 2 arcsec pixel scale
(see Stevens et al. 2004 for more details). These maps were then convolved with
a Gaussian of full width half maximum 4 pixels (850~$\mu$m) or 3 pixels
(450~$\mu$m) giving final resolutions of $14.2$ (850~$\mu$m) and $8.6$ arcsecs
(450~$\mu$m). The convolved maps with signal-to-noise (S/N) contours are shown
in Fig.\ \ref{fig:allmaps}.

\begin{table}
\footnotesize
\centering
\def\baselinestretch{1}                                     
\caption[dum]{\small Observing log of the SCUBA mapping observations. For each
field we give the date of observation, the number of 64-pt jiggle maps where 10
maps is equivalent to $\sim20$\ mins of integration, and the opacity range
($\tau_{\rm wvm}$) as given at $225$\ GHz from measurements with the JCMT water
vapour radiometer. Observations of RX~J$094144.51+385434.8$ were published by
Stevens et al. (2004).}
\label{table:log}
\vspace*{0.1in}
\begin{tabular}{llcc}\hline
 Name & Date & $N_{\rm map}$ & $\tau_{\rm wvm}$ \\ \hline
RX\ J$005734.78-272827.4$ & 20031008 & 20 & $0.07-0.08$ \\
 & 20031209 & 40 & $0.06-0.08$\\
 & 20031212 & 20 & $0.03-0.04$\\
 & 20031215 & 10 & $0.04-0.05$\\
 & 20031226 & 20 & $0.04-0.05$\\
 & 20040108 & 20 & $0.04-0.05$\\
 & 20040109 & 10 & $0.05-0.06$\\
 & 20040826 & 20 & $0.06-0.08$\\
 & 20040829 & 20 & $0.07-0.08$\\
 & 20041230 & 20 & $0.04-0.05$\\
RX\ J$121803.82+470854.6$ & 20041218 & 40 & $0.07-0.08$\\
 & 20041229 & 53 & $0.05-0.06$\\
 & 20041230 & 100 & $0.03-0.08$ \\
RX\ J$124913.86-055906.2$ & 20031224 & 20 & $0.04-0.05$\\
 & 20031226 & 40 & $0.04-0.05$\\
 & 20040107 & 60 & $0.04-0.06$\\
 & 20040108 & 60 & $0.04-0.05$\\
 & 20040109 & 60 & $0.05-0.06$\\
 & 20040110 & 40 & $0.05-0.06$\\
RX\ J$163308.57+570258.7$ & 20030708 & 80 & $0.06-0.08$\\
 & 20030709 & 80 & $0.08-0.09$\\
 & 20030711 & 80 & $0.07-0.08$\\
 & 20030712 & 60 & $0.05-0.07$\\
 & 20040109 & 20 & $0.04-0.05$\\
 & 20040110 & 60 & $0.05-0.06$\\
 & 20040111 & 20 & $0.05-0.06$\\ \hline
\end{tabular}
\end{table}

\subsection{{\em Spitzer\/} observations}

We observed each field with the IRAC (at 4.5 and 8.0 $\mu$m) and MIPS (at 24
$\mu$m) cameras on-board the {\em Spitzer Space Telescope}. IRAC observations
were made between 2005 March and 2005 July while the MIPS observations were
made between 2005 March and 2005 December.  For the IRAC observations we used
5-pt Gaussian dithers with frame times of 100 or 200s depending on the
predicted brightness of the target. Frame times for the MIPS observations were
between 3 and 10~s. These observations are summarised in
Table~\ref{table:spitlog}.

The IRAC frames were processed by the {\em Spitzer\/} Science Center using the
standard pipeline version S14.0.0.
The images were then cropped to only include the region of constant exposure
time; this allows catalogues to be constructed from regions with similar noise
properties.  Source extraction was performed with {\sc SExtractor} (Bertin \&
Arnouts 1996) with a fixed aperture of $3.5^{\prime\prime}$ and a local
background fit. Aperture corrections of $1.6$ and $2.13$ were applied to the
4.5 and 8.0~$\mu$m flux densities respectively, in accord with the values
determined from the {\em Spitzer\/} First Look Survey (Lacy et al. 2005).

For MIPS observations, the $24~\mu$m BCD images were fitted with a plane after
first masking out the bright objects.  This plane was then
subtracted from each BCD frame resulting in a flat background image. These
images were then combined with the {\sc IRAF} task {\sc imcombine} using the
astrometric solution from each individual BCD image. This procedure resulted in
a much improved 24~$\mu$m image, free of the jail-bar pattern common to the
PBCD images supplied from the {\em Spitzer\/} pipeline. Again, only the central
$3\times 3$~arcmin$^2$ region of the images was used to detect sources. We used
{\sc SExtractor} to find sources and the {\sc phot} script in IDL to measure
their flux density within a $13$~arcsec radius aperture.  All flux densities
were then subjected to an aperture correction of 1.167 (see Seymour et
al. 2007).

\begin{table*}
\footnotesize
\centering
\def\baselinestretch{1}                                     
\caption[dum]{\small Observing log of the {\em Spitzer\/} observations. For each
field we give the date of observation and the on-source integration time ($T_{\rm
int}$) for each instrument.}
\label{table:spitlog}
\vspace*{0.1in}
\begin{tabular}{llccc}\hline
Name & \multicolumn{2}{c}{IRAC 4.5/8.0 $\mu$m} & \multicolumn{2}{c}{MIPS 24 $\mu$m} \\ 
& Date & $T_{\rm int}$ (s) & Date & $T_{\rm int}$ (s) \\ \hline
RX\ J$005734.78-272827.4$ & 20050717 & $2000$ & 20051202 & $432$ \\
RX\ J$094144.51+385434.8$ & 20050506 & $3000$ & 20050412 & $336$ \\
RX\ J$121803.82+470854.6$ & 20050506 & $2000$ & 20050519 & $210$ \\
RX\ J$124913.86-055906.2$ & 20050715 & $5000$ & 20050627 & $636$ \\
RX\ J$163308.57+570258.7$ & 20050326 & $1000$ & 20050306 & $970$ \\ \hline
\end{tabular}
\end{table*}

\section{The submillimetre catalogues}

Our philosophy is to match our analysis as closely as possible to that
performed by the SHADES consortium (Coppin et al. 2006) thus allowing a direct
comparison of the number counts determined for blank fields and those in the
highly biased regions of the Universe surrounding our QSOs. Most of the
analysis in this section pertains to the 850~$\mu$m data. For the 450~$\mu$m
data we extract a catalogue at a fixed S/N threshold but do not compute number
counts.

\subsection{Source extraction}

\label{section:se}

We adopt the source extraction algorithm described by Scott et al. (2002) which
performs a maximum-likelihood fit to the flux densities of all potentially
significant peaks identified on the convolved maps. For each source we took the
convolved S/N map and searched for all peaks with S/N$>1.5$. A peak is found if
a $10\times10$ arcsec grid centred on any pixel fails to find another pixel in
the grid with higher S/N. In practice, this means that real sources separated
by less than a beam width can be missed by the extraction algorithm. This is
the case for RX~J$094144$ where we fail to recover two sources which
are clearly real because we see them in the higher resolution $450~\mu$m data
but are blended together at $850~\mu$m.  We then took the unconvolved signal
map and convolved it with appropriately scaled beam maps centred at these
locations. The amplitudes of these beam maps were then adjusted until the
overall $\chi^2$ of the fit was minimised. For two sources with azimuth
chop/nod schemes the beam was reconstructed by going back to the observing logs
and calculating the position angle of the chop throw on the sky at the
mid-point of each observation. We then added these together, weighted by the
integration time for each observation. 

This method is suitable for extracting sources in fields where confusion is
likely to be a problem; it also has the advantage of using the off-positions to
distinguish between real sources and noise spikes, and increases the S/N by
considering the flux density in the off-positions. The method was used for all
four independent SHADES reductions although note that `Reduction A' of Coppin
et al. (2006) searched for peaks on the convolved signal maps rather than the
convolved S/N maps. We tried both methods but found that peaks towards the map
edges were better recovered from the S/N maps.

Uncertainties in submillimetre positions measured on SCUBA maps have been
discussed extensively by Ivison et al. (2007) who show that the 1-$\sigma$
error circle is given by $0.9\,\theta_{\rm A}({\rm SNR}^2-3)^{-1/2}$ where
$\theta_{\rm A}$ is the FWHM of the beam and SNR is the raw map signal-to-noise
ratio. A typical source with SNR$\sim3-4$ at $850~\mu$m would thus have a
1-$\sigma$ positional accuracy of $3-5$ arcsec.

\subsection{Flux density deboosting}

A well known phenomenon common to all deep submillimetre surveys is that of
flux density boosting. Real sources falling in regions of positive noise will
have their intrinsic flux densities boosted, raising the likelihood of
inclusion in the catalogue. This effect becomes increasingly significant as the
intrinsic flux density of the source approaches the limit of the survey. If not
corrected for, flux density boosting will therefore populate the survey
catalogue with an excess of low flux-density sources.

We use the Bayesian flux density deboosting recipe of Coppin et al. (2005,
2006) which calculates a posterior flux density probability distribution for
each source assuming a prior distribution of real flux densities constructed
from existing data. The prior is constructed using artificial skies which are
sampled with the same observing technique used for the real data. Since we used
two different chop/nod schemes we employ two priors; a two position chop scheme
for the 3 fields observed with East-West chop throws and the SHADES observing
pattern for the two fields that were chopped in azimuth. SHADES employed a 6
position chop/nod scheme, chopping in both right ascension and declination with
chop throws of 30, 44 and 68 arcsecs.  We note that while this method is not a
true match to an azimuth chop scheme, where the chop position angle rotates on
the sky throughout the observation, it is nevertheless a better approximation
than a two chop scheme. In any case, running the deboosting algorithm with
either simulated observing technique resulted in the same catalogue give or
take 1 or 2 sources close to the selection threshold. Deboosted flux densities
are listed in Table~\ref{table:850cat}.

\subsection{Catalogue membership}

\label{section:catmem}

\begin{figure*}
\setlength{\unitlength}{1in}
\begin{picture}(7.0,8.3)
\includegraphics{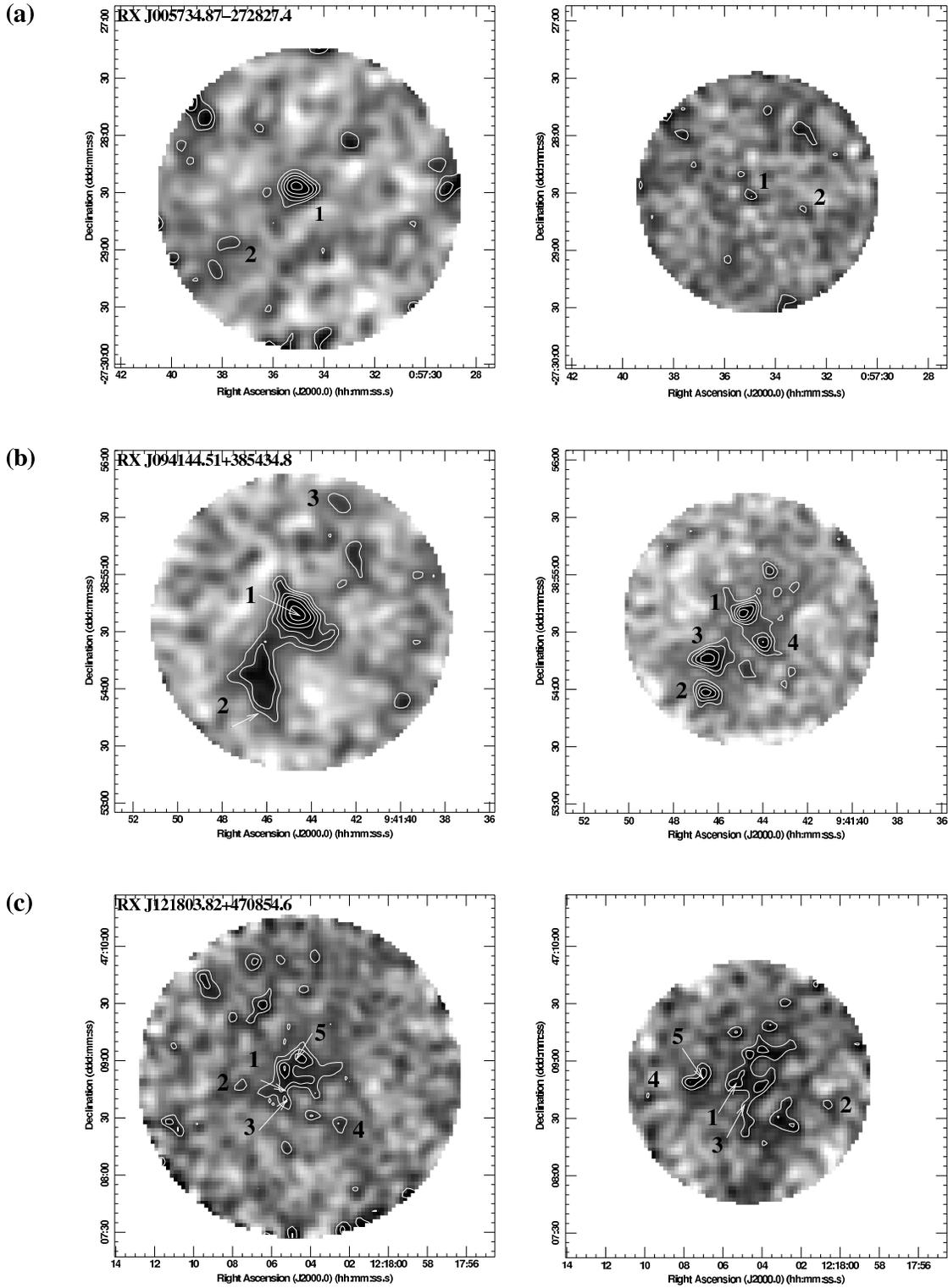}
\end{picture}
 \caption[dum]{SCUBA maps of (a) RX\,J$005734.87-272827.4$, (b) RX\,J$094144.51+385434.8$, (c) RX\,J$121803.82+470854.6$, (d) RX\,J$124913.86-055906.2$ and (e) RX\,J$163308.57+570256.7$ at 850\ (left) and 450\ $\mu$m (right). The grey-scales show flux density while the contours, 
which show signal-to-noise, are calculated using the noise maps (see
 text). Contour levels are 2, 3, 4, 5 and 6\ $\sigma$ except for RX\,J$124913.86-055906.2$ where they are 2, 3, 4, 5, 6, 8, 10 and 12\ $\sigma$. These contours refer to the significance levels on the pre-extracted maps as shown. Note that real sources will have higher significance when extracted with the beam profile because emission in the off-positons is also factored into the total signal-to-noise.}
\label{fig:allmaps}
\end{figure*}
\begin{figure*}
\setlength{\unitlength}{1in}

\addtocounter{figure}{-1}

\begin{picture}(7.0,5.5)
\includegraphics{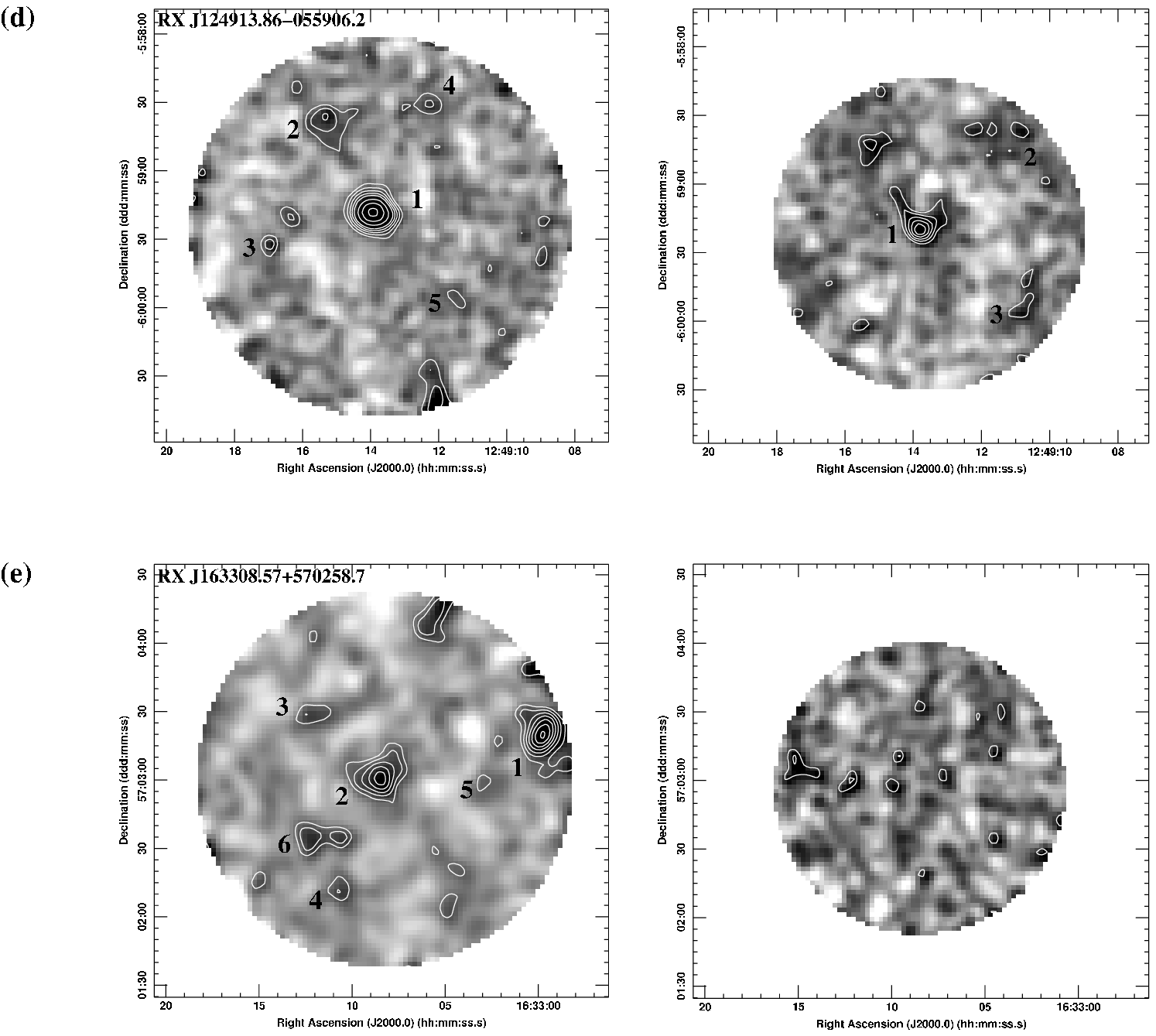}
\end{picture}
\caption[dum]{- continued}
\end{figure*}

\begin{figure}
\setlength{\unitlength}{1in}
\begin{picture}(3.5,3.2)
\includegraphics{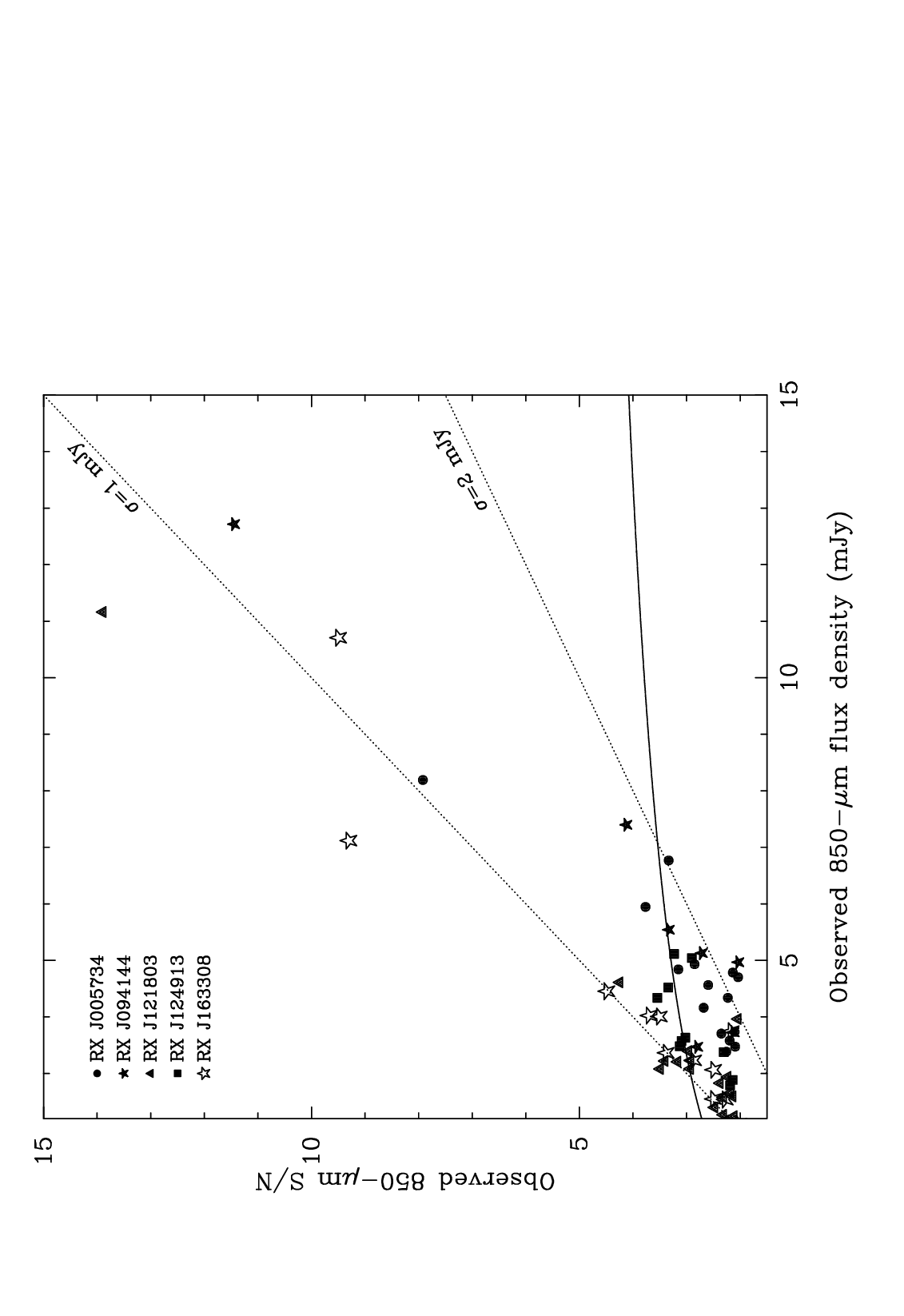}
\end{picture}
\caption[dum]{Extracted 850\ $\mu$m flux density against extracted 850\ $\mu$m
signal-to-noise (uncorrected for flux density boosting). All sources
extracted with S/N$>2$ are plotted; those that fall
above the solid line have $>95$ per cent posterior probability that their
flux density is $>0$ mJy. These sources make it into our catalogue. Dotted
lines show noise levels of 1 and 2 mJy/beam. }
\label{fig:catalogue}
\end{figure}

As a first cut we took all sources found by the extraction algorithm with a
S/N$>2.0$. This catalogue was then trimmed to include only those sources which
have less than $5$ per cent of their posterior probability distribution below
$0$ mJy. This cut is exactly that used by Coppin et al. (2006) for the SHADES
catalogue. In Fig.~\ref{fig:catalogue} we plot extracted 850-$\mu$m S/N against
extracted 850-$\mu$m flux density. Those sources falling above the solid line,
and that remain there after deboosting, meet the above criterion and are
included in the final catalogue (Table~\ref{table:850cat}). In the 5 fields we
find 21 sources at 850~$\mu$m including 4 of the 5 QSOs.  At 450~$\mu$m we
simply list those sources extracted with S/N$>3$ (Table~\ref{table:450cat}) and
do not attempt to deboost their flux densities. We assume that an 850~$\mu$m
source is also detected at 450~$\mu$m if their extracted positions are within
10 arcsec. Using this criterion we recover 8 of the sources detected at
850~$\mu$m including 3 of the QSOs. A further 6 sources detected at 450~$\mu$m
are not detected at 850~$\mu$m. Two of these objects lie in the RX~J$094144$
field and were missed at 850~$\mu$m due to blending as discussed in
Section~\ref{section:se}. We note that the QSO RX~J$121803.82+470854.6$ is not
formally detected at either 850 or 450~$\mu$m even though it was detected by
Page et al. (2001) using SCUBA in photometry mode. This QSO is discussed
further in Appendix A. 

\subsection{Catalogue completeness}

Extensive simulations were performed to determine the completeness as a function of flux density. For each field we added false sources to the signal maps by scaling the normalised beam map to have an amplitude of 2, 3, 4, 5, 6, 7, 8, 9, 10, 12, 14, 16 and 18 mJy. We then attempted to recover each source with the extraction algorithm and repeated this procedure 1000 times per field. We consider a source detected if its recovered flux density is within a factor of 2 of its input flux density, if its recovered position is within 7 arcsecs of its input position and if it makes it into our catalogue as described in the previous section. These criteria are those adopted by the SHADES survey. We plot the fractional completeness for each field as a function of input flux density in Fig.~\ref{fig:comp}. The fits to the data have the form $Completeness=(S^a)/(b+cS^a)$ where $S$ is the 850-$\mu$m flux density and $a,b,c$ are free parameters.  It is apparent that the completeness is a function of both the average noise level of the maps and the structure that they contain. We note that the fractional completeness converges to $\sim0.85-0.9$ rather than 1 at high flux densities. This is due to the failure of the extraction algorithm to recover input sources which fall very close to the field edges. The effect will be worse towards the top and bottom centres of the fields where both off-positions fall outside the area used for extraction. We confirmed that this is the case by performing an 18~mJy simulation over the centre region of one of the maps. This simulation returned a fractional completeness of 1.0. However, since real sources may also fall on the map edges in exactly the same way that randomly generated false sources do, we do not correct for this edge effect. One other feature of these simulations worthy of remark is that the shape of the completeness curve appears to be sensitive to the adopted chopping scheme. Those fields observed with an azimuthal chop/nod scheme have relatively high completeness at low flux densities, presumably due to an absence of the deep negative off-positions that are produced by chopping at a fixed position on the sky (most fields have a bright QSO at their centre), but they are relatively less complete at high flux densities where the absence of these off-positions becomes more of a hindrance.

\begin{table*}
\footnotesize
\centering
\def\baselinestretch{1}                                     
\caption[dum]{\small The 850-$\mu$m catalogue. Coordinates give the extracted
  position of the source. Also listed are extracted flux density and S/N
  together with deeboosted flux density (see text for details). Quoted
  uncertainties on the raw flux densities include contributions from
  the calibration uncertainty and the extracted signal-to-noise values while
  those on the deboosted flux densities are purely statistical.}
\label{table:850cat}
\vspace*{0.1in}
\begin{tabular}{lccccrcl}\hline
Field & ID\# & RA & Dec & $S_{850}$ & $S_{850}$ (deb.) & Map S/N & Note\\ 
       & &(J2000.0) & (J2000.0)& (mJy/beam) & (mJy/beam) & & \\ \hline
RX\,J$005734.78-272827.4$ & $1$ & $00^{\rm h}\ 57^{\rm m}\ 35^{\rm s}.21$ & $-27^{\circ}\
       28^{\prime}\ 27^{\prime\prime}.9$ & $8.2\pm1.2$ & $7.8^{+1.0}_{-0.9}$ &
       $7.9$ & QSO \\ [5pt]
& $2$ & $00^{\rm h}\ 57^{\rm m}\ 37^{\rm s}.79$ & $-27^{\circ}\ 28^{\prime}\
       57^{\prime\prime}.9$ & $5.9\pm1.6$ & $4.3^{+1.7}_{-2.1}$ & $3.8$ & \\ [5pt] 

RX\,J$094144.51+385434.8$ & $1$ & $09^{\rm h}\ 41^{\rm m}\ 44^{\rm s}.84$ & $+38^{\circ}\
       54^{\prime}\ 36^{\prime\prime}.9$ & $12.7\pm1.4$ & $12.4^{+1.1}_{-1.0}$
       & $11.4$ & QSO \\ [5pt]
& $2$ & $09^{\rm h}\ 41^{\rm m}\ 46^{\rm s}.53$ & $+38^{\circ}\
       53^{\prime}\ 54^{\prime\prime}.7$ & $7.4\pm1.9$ & $5.7^{+2.0}_{-2.1}$ &
       $4.1$ & \\ [5pt]
& $3$ & $09^{\rm h}\ 41^{\rm m}\ 42^{\rm s}.80$ & $+38^{\circ}\
       55^{\prime}\ 34^{\prime\prime}.8$ & $5.5\pm1.7$ & $3.1^{+2.1}_{-2.2}$ &
       $3.3$ &\\ [5pt] 

RX\,J$121803.82+470854.6$ & $1$ & $12^{\rm h}\ 18^{\rm m}\ 05^{\rm s}.51$ & $+47^{\circ}\
       08^{\prime}\ 55^{\prime\prime}.1$ & $4.3\pm1.3$ & $3.1^{+1.3}_{-1.6}$ &
       $3.5$  & \\ [5pt]
& $2$ & $12^{\rm h}\ 18^{\rm m}\ 07^{\rm s}.66$ & $+47^{\circ}\
       08^{\prime}\ 45^{\prime\prime}.1$ & $4.5\pm1.4$ & $2.8^{+1.6}_{-1.7}$ &
       $3.3$ & \\ [5pt]
& $3$ & $12^{\rm h}\ 18^{\rm m}\ 05^{\rm s}.53$ & $+47^{\circ}\
       08^{\prime}\ 36^{\prime\prime}.9$ & $3.6\pm1.2$ & $2.1^{+1.3}_{-1.3}$ &
       $3.1$ & \\ [5pt]
& $4$ & $12^{\rm h}\ 18^{\rm m}\ 02^{\rm s}.76$ & $+47^{\circ}\
       08^{\prime}\ 24^{\prime\prime}.9$ & $3.5\pm1.2$ & $2.1^{+1.3}_{-1.2}$ &
       $3.1$ & \\ [5pt]
& $5$ & $12^{\rm h}\ 18^{\rm m}\ 04^{\rm s}.74$ & $+47^{\circ}\
       08^{\prime}\ 59^{\prime\prime}.0$ & $3.6\pm1.2$ & $1.9^{+1.5}_{-1.2}$ &
       $3.0$ & \\ [5pt]  

RX\,J$124913.86-055906.2$ & $1$ & $12^{\rm h}\ 49^{\rm m}\ 13^{\rm s}.99$ & $-05^{\circ}\
       59^{\prime}\ 19^{\prime\prime}.3$ & $11.2\pm1.1$ & $11.0^{+0.7}_{-0.7}$
       & $13.9$  & QSO\\ [5pt]
& $2$ & $12^{\rm h}\ 49^{\rm m}\ 15^{\rm s}.45$ & $-05^{\circ}\
       58^{\prime}\ 37^{\prime\prime}.3$ & $4.6\pm1.1$ & $3.8^{+1.1}_{-1.1}$ &
       $4.3$ & \\ [5pt]
& $3$ & $12^{\rm h}\ 49^{\rm m}\ 17^{\rm s}.06$ & $-05^{\circ}\
       59^{\prime}\ 33^{\prime\prime}.3$ & $3.1\pm0.9$ & $2.3^{+0.9}_{-1.0}$ &
       $3.5$ & \\ [5pt]
& $4$ & $12^{\rm h}\ 49^{\rm m}\ 12^{\rm s}.37$ & $-05^{\circ}\
       58^{\prime}\ 33^{\prime\prime}.5$ & $3.2\pm1.0$ & $2.3^{+1.0}_{-1.0}$ &
       $3.4$ & \\ [5pt]
& $5$ & $12^{\rm h}\ 49^{\rm m}\ 11^{\rm s}.56$ & $-05^{\circ}\
       59^{\prime}\ 59^{\prime\prime}.3$ & $3.2\pm1.0$ & $2.1^{+1.1}_{-1.1}$ &
       $3.2$ & \\ [5pt]

RX\,J$163308.57+570258.7$ & $1$ & $16^{\rm h}\ 32^{\rm m}\ 59^{\rm s}.99$ & $+57^{\circ}\
       03^{\prime}\ 18^{\prime\prime}.7$ & $10.7\pm1.4$ & $10.4^{+1.0}_{-1.1}$
       &  $9.5$ & \\ [5pt]
& $2$ & $16^{\rm h}\ 33^{\rm m}\ 08^{\rm s}.82$ & $+57^{\circ}\
       02^{\prime}\ 58^{\prime\prime}.7$ & $7.1\pm0.9$ & $6.9^{+0.7}_{-0.7}$ &
       $9.3$ & QSO \\ [5pt]
& $3$ & $16^{\rm h}\ 33^{\rm m}\ 12^{\rm s}.74$ & $+57^{\circ}\
       03^{\prime}\ 26^{\prime\prime}.7$ & $4.5\pm1.0$ & $3.8^{+1.0}_{-1.0}$ &
       $4.5$ & \\ [5pt]
& $4$ & $16^{\rm h}\ 33^{\rm m}\ 10^{\rm s}.78$ & $+57^{\circ}\
       02^{\prime}\ 10^{\prime\prime}.7$ & $4.0\pm1.1$ & $3.0^{+1.2}_{-1.3}$ &
       $3.7$ & \\ [5pt]
& $5$ & $16^{\rm h}\ 33^{\rm m}\ 03^{\rm s}.18$ & $+57^{\circ}\
       02^{\prime}\ 58^{\prime\prime}.7$ & $4.0\pm1.2$ & $2.8^{+1.3}_{-1.4}$ &
       $3.5$ & \\ [5pt]
& $6$ & $16^{\rm h}\ 33^{\rm m}\ 12^{\rm s}.61$ & $+57^{\circ}\
       02^{\prime}\ 32^{\prime\prime}.0$ & $3.4\pm1.1$ & $2.3^{+1.1}_{-1.1}$ &
       $3.4$ & \\
\hline
\end{tabular}
\end{table*}

\begin{table*}
\footnotesize
\centering
\def\baselinestretch{1}                                     
\caption[dum]{\small The 450-$\mu$m catalogue. Coordinates give the extracted
  position of the source. Also listed are extracted flux density and
  S/N. Quoted uncertainties on the raw flux densities include contributions
  from the calibration uncertainty and the extracted signal-to-noise values.
  The last column gives the likely ID if the 450~$\mu$m position falls within
  10 arcsecs of either the QSO or one of the objects extracted at 850~$\mu$m.}
\label{table:450cat}
\vspace*{0.1in}
\begin{tabular}{lcccccl}\hline
Field & ID\# & RA & Dec & $S_{450}$ & Map S/N & Note\\ 
       & &(J2000.0) & (J2000.0)& (mJy/beam) & &\\ \hline
RX\,J$005734.78-272827.4$ & $1$ & $00^{\rm h}\ 57^{\rm m}\ 35^{\rm s}.08$ & $-27^{\circ}\
       28^{\prime}\ 34^{\prime\prime}.1$ & $32.4\pm10.2$ & $4.1$ & QSO \\ [1pt]
& $2$ & $00^{\rm h}\ 57^{\rm m}\ 33^{\rm s}.13$ & $-27^{\circ}\ 28^{\prime}\
       40^{\prime\prime}.1$ & $25.9\pm9.6$ & $3.2$ &  \\ [5pt] 

RX\,J$094144.51+385434.8$ & $1$ & $09^{\rm h}\ 41^{\rm m}\ 45^{\rm s}.06$ & $+38^{\circ}\
       54^{\prime}\ 36^{\prime\prime}.7$ & $44.7\pm10.2$ & $9.1$ & QSO \\ [1pt]
& $2$ & $09^{\rm h}\ 41^{\rm m}\ 46^{\rm s}.73$ & $+38^{\circ}\
       53^{\prime}\ 56^{\prime\prime}.7$ & $41.2\pm10.0$ & $7.3$ & 850\#2  \\ [1pt]
& $3$ & $09^{\rm h}\ 41^{\rm m}\ 46^{\rm s}.73$ & $+38^{\circ}\
       54^{\prime}\ 14^{\prime\prime}.9$ & $33.5\pm8.8$ & $5.9$ & \\ [1pt]
& $4$ & $09^{\rm h}\ 41^{\rm m}\ 44^{\rm s}.16$ & $+38^{\circ}\
       54^{\prime}\ 22^{\prime\prime}.7$ & $28.8\pm7.8$ & $5.5$ & \\ [5pt]

RX\,J$121803.82+470854.6$ & $1$ & $12^{\rm h}\ 18^{\rm m}\ 05^{\rm s}.51$ & $+47^{\circ}\
       08^{\prime}\ 47^{\prime\prime}.1$ & $35.1\pm8.9$ & $6.4$ & 850\#1 \\ [1pt]
& $2$ & $12^{\rm h}\ 18^{\rm m}\ 00^{\rm s}.80$ & $+47^{\circ}\
       08^{\prime}\ 34^{\prime\prime}.9$ & $31.9\pm10.7$ & $3.7$ & \\ [1pt]
& $3$ & $12^{\rm h}\ 18^{\rm m}\ 05^{\rm s}.11$ & $+47^{\circ}\
       08^{\prime}\ 35^{\prime\prime}.1$ & $30.4\pm11.0$ & $3.3$ & 850\#3 \\ [1pt]
& $4$ & $12^{\rm h}\ 18^{\rm m}\ 10^{\rm s}.02$ & $+47^{\circ}\
       08^{\prime}\ 41^{\prime\prime}.1$ & $24.3\pm8.8$ & $3.3$ & \\ [1pt]
& $5$ & $12^{\rm h}\ 18^{\rm m}\ 07^{\rm s}.08$ & $+47^{\circ}\
       08^{\prime}\ 52^{\prime\prime}.9$ & $21.9\pm8.3$ & $3.1$ & 850\#2\\ [5pt]

RX\,J$124913.86-055906.2$ & $1$ & $12^{\rm h}\ 49^{\rm m}\ 13^{\rm s}.97$ & $-05^{\circ}\
       59^{\prime}\ 21^{\prime\prime}.3$ & $34.9\pm8.1$ & $8.7$ & QSO\\ [1pt]
& $2$ & $12^{\rm h}\ 49^{\rm m}\ 11^{\rm s}.03$ & $-05^{\circ}\
       58^{\prime}\ 37^{\prime\prime}.5$ & $21.6\pm7.7$ & $3.4$ & \\ [1pt]
& $3$ & $12^{\rm h}\ 49^{\rm m}\ 11^{\rm s}.03$ & $-05^{\circ}\
       59^{\prime}\ 57^{\prime\prime}.5$ & $33.7\pm12.5$ & $3.2$ & 850\#5 \\ [1pt]
\hline
\end{tabular}
\end{table*}

\section{Integral number counts}

\begin{figure*}
\setlength{\unitlength}{1in}
\begin{picture}(8.0,3.0)
\includegraphics{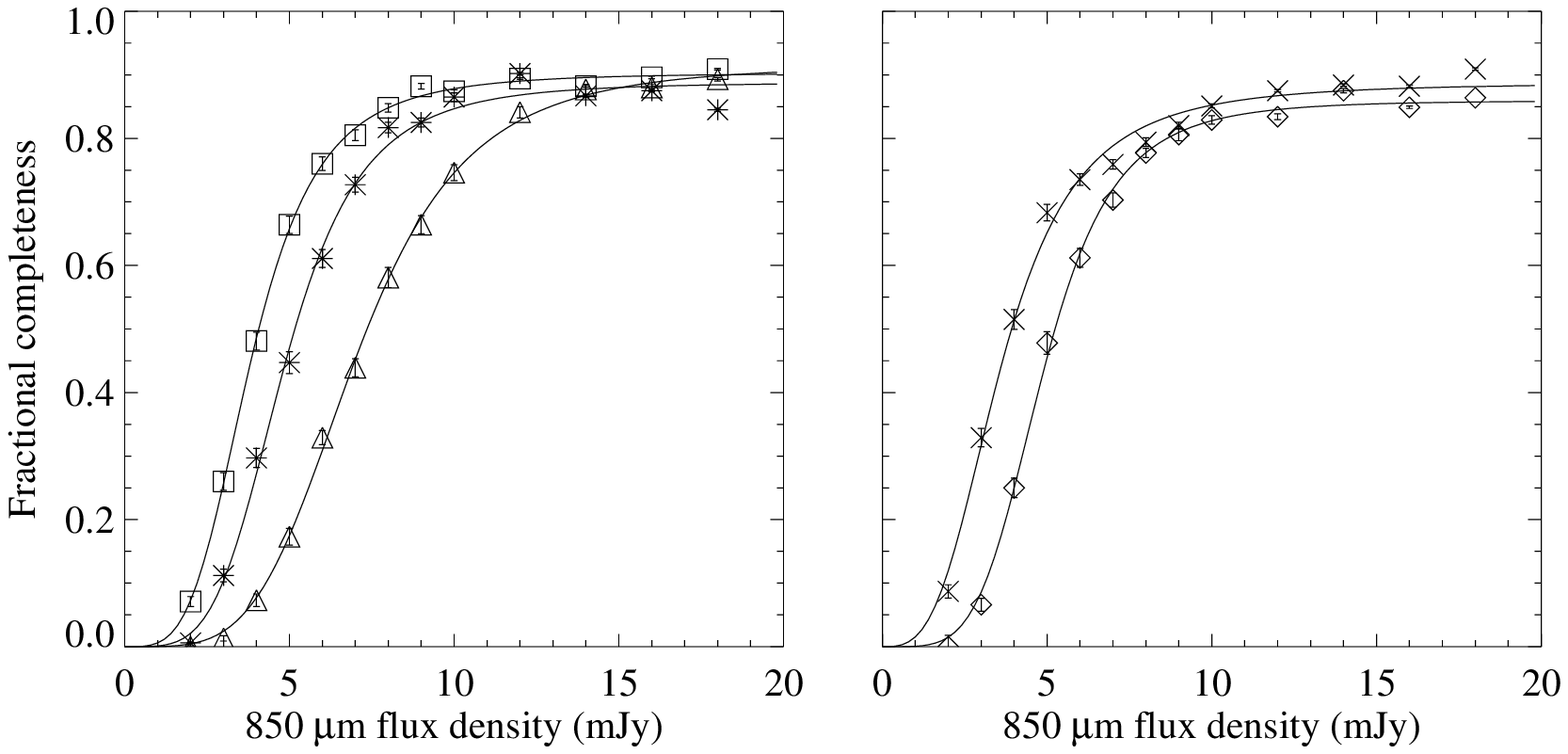}
\end{picture}
\caption[dum]{Fractional completeness as a function of 850-$\mu$m flux density
for each of the five fields. The curves are fits to the data (see text for
details). From left-hand panel shows the completeness for RX\ J$163308$
(squares), RX\ J$121803$ (diamonds) and RX\ J$094144$ (triangles) while the
right-hand panel shows the completeness for RX\ J$124913$ (asterisks) and RX\
J$005734$ (diamonds).}
\label{fig:comp}
\end{figure*}

Number counts at 850 $\mu$m were calculated from the deboosted flux densities after first making a correction for the completeness of the catalogue for each field using the curves shown in Fig.~\ref{fig:comp}. The total area of the survey is 28.4 arcmin$^2$. We follow Coppin et al. (2006) so that our counts are calculated in the same manner as the SHADES counts. The method uses the deboosted flux density probability distributions as input for a modified boot-strapping simulation which estimates the number counts and their errors. This method allows the flux density of an individual source to be shared appropriately between adjacent bins. We quote counts in bins of width 2~mJy (QSOs included; Table~\ref{table:counts}) and 1~mJy (QSOs removed; Table~\ref{table:counts2}). These counts represent the mean value in each bin with the quoted error being the frequentist 68\% confidence interval in each flux bin based on boot-strapped Monte Carlo simulations, including Poisson counting errors.  While including the QSOs gives a biased representation of the over-density in these fields, this method gives the total number of submillimetre detected galaxies. This approach is also analogous to that adopted by previous workers on QSO over-densities (e.g. Priddey et al. 2008), although we stress that the counts are only appropriate for X-ray absorbed QSOs that are pre-selected to be submillimetre luminous.

Quoted number counts exclude the contribution from sources 3, 4 and 5 in the RX~J$121803$ field. All three have deboosted flux densities close to 2~mJy where the completeness estimate from our simulations is very low ($<1$ per cent) and therefore subject to a large uncertainty. The implication is that number counts in the lowest flux density bins will be underestimated. If we assume that the true completeness of this field at 2 mJy is 5 per cent (a reasonable value from inspection of Fig.~\ref{fig:comp}) and that the three sources contribute entirely to the lowest flux density bin then the number counts in this bin would increase by $~70$ per cent. However, no correction is applied to the quoted counts.

\begin{figure}
\setlength{\unitlength}{1in}
\begin{picture}(7.0,3.0)
\includegraphics{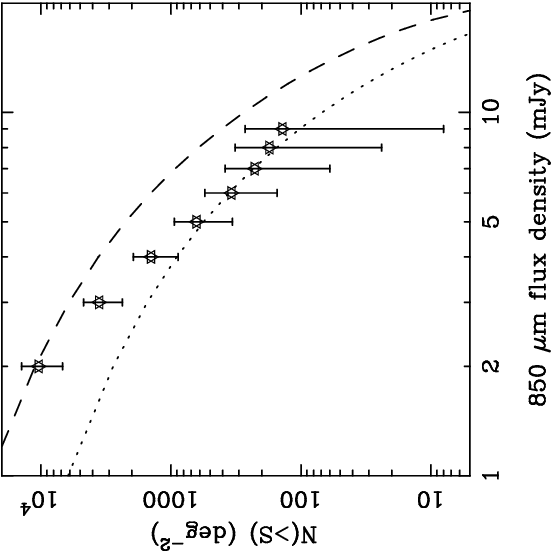}
\end{picture}
\caption[dum]{Integral number counts plotted in 1 mJy bins. The QSOs have been removed from the counts. The dotted line is a best-fit Schechter function to the SHADES counts as described in Coppin et al. (2006) while the dashed line is the result of multiplying this fit by 4 and is shown for reference.}
\label{fig:counts}
\end{figure}

The counts (QSOs removed) are plotted in Fig.~\ref{fig:counts} which also shows number counts determined for the SHADES survey (Coppin et al. 2006). We note that counts determined at 870~$\mu$m from a slightly larger survey conducted with the Large APEX Bolometer Camera (LABOCA) are consistent with those measured for SHADES at flux densities below 3~mJy and are about a factor of 2 below the SHADES counts at higher flux densities (Wei\ss{} et al. 2009). The dashed line is not a fit to the data but is simply the best-fit Schechter function to the SHADES counts multiplied by 4. From this plot we can see that the observed over-density of sources around the QSOs arises almost entirely from galaxies with deboosted flux densities in the range $\sim2-4$~mJy. At higher flux densities the counts in the QSO fields are matched very well to those determined from SHADES data for the field.

\begin{table}
\footnotesize
\centering
\caption[dum]{\small Combined integral number counts for the 5 QSO fields including the QSOs. Bins
are quoted at the lower flux density bound. The bins are of width 2 mJy. SHADES counts are taken from
Coppin et al. (2006).}
\label{table:counts}
\vspace*{0.1in}
\begin{tabular}{ccc}\hline
Flux density & \multicolumn{2}{c}{850 $\mu$m integral counts, $N(>S)$}\\ 
(mJy) &  \multicolumn{2}{c}{(deg$^{-2})$} \\ 
& This work & SHADES \\\hline
$2.0$ & $11072^{+6019}_{-2791}$& $2506^{+407}_{-407}$ \\ [5pt]
$4.0$ & $2074^{+1060}_{-470}$& $844^{+117}_{-117}$\\ [5pt] 
$6.0$ & $953^{+507}_{-305}$& $362^{+61}_{-59}$\\ [5pt]
$8.0$ & $552^{+278}_{-258}$& $150^{+35}_{-35}$\\ [5pt]
$10.0$ & $375^{+190}_{-237}$& $68^{+21}_{-21}$\\ [5pt]
$12.0$ & $120^{+85}_{-105}$& $33^{+16}_{-15}$\\ \hline
\end{tabular}
\end{table}

\begin{table}
\footnotesize
\centering
\caption[dum]{\small Combined integral number counts for the 5 QSO fields excluding the QSOs. Bins
are quoted at the lower flux density bound. The bins are of width 1 mJy.}
\label{table:counts2}
\vspace*{0.1in}
\begin{tabular}{cc}\hline
Flux density & 850 $\mu$m integral counts, $N(>S)$\\ 
(mJy) & (deg$^{-2})$ \\ 
& This work \\\hline
$2.0$ & $10420^{+3676}_{-3584}$ \\ [5pt]
$3.0$ & $3574^{+1129}_{-1211}$ \\ [5pt] 
$4.0$ & $1423^{+524}_{-540}$\\ [5pt]
$5.0$ & $636^{+308}_{-300}$\\ [5pt]
$6.0$ & $343^{+205}_{-190}$ \\ [5pt]
$7.0$ & $227^{+156}_{-167}$ \\ [5pt]
$8.0$ & $175^{+146}_{-151}$ \\ [5pt]
$9.0$ & $139^{+130}_{-131}$ \\\hline
\end{tabular}
\end{table}

\section{Mid-infrared counterparts}

Likely mid-infrared counterparts to the submillimetre galaxies are given in
Table~\ref{table:spit}. For each field we searched within fixed radii of the
extracted submillimetre positions. These radii of 5 arcsec at $4.5$ and
$8.0~\mu$m, and 10 arcsec at $24~\mu$m were chosen to approximately match the
theoretical error circles on the submillimetre positions added in quadrature
with the uncertainty on the mid-infrared positions, which are significant at
$24~\mu$m (see Ivison et al. 2007 for a full discussion). The corrected $P$
statistic (Downes et al. 1986) is given for each source found within the search
radii. This quantity is given by $P=1-{\rm exp}\,[-P_*\{1+{\rm ln}\,(P_{\rm
c}/P_*)\}]$ where $P_c=\pi r_{\rm s}^2 N_{\rm T}$, $P_*=\pi r^2 N$, $r_{\rm s}$
is the search radius, $N_{\rm T}$ is the surface density of objects at the
limiting flux density of the catalogue, $N$ is the surface density of objects
brighter than $S$ where $S$ is the flux density of a potential counterpart
galaxy located a distance, $r$, away from the submillimetre position.

A value of $P\leq0.05$ is typically used as a threshold to select robust radio
IDs (e.g. Ivison et al. 2002, 2007) to submillimetre galaxies. 
We select all mid-infrared
sources found within our search radii as potential counterparts and use the $P$
statistic to distinguish between competing IDs when more than one galaxy is
found. We note that almost all SMGs listed in Table~\ref{table:spit} have
mid-infrared counterparts for which the computed $P$ value is $<0.05$ in at
least one waveband, and can therefore be considered robust based on this
criterion.  

The 4 QSOs detected at submillimetre wavelengths have counterparts in all 3
{\em Spitzer\/} bands. In addition, we find counterparts in at least one
waveband for $\sim 60$ per cent of the SMGs.  Postage stamp images of the
mid-infrared counterparts are shown in Figs~\ref{fig:post1} and \ref{fig:post2}
which can be found in Appendix A.

\begin{table*}
\footnotesize
\centering
\def\baselinestretch{1}                                     
\caption[dum]{\small {Candidate \em Spitzer\/} IRAC and MIPS counterparts to
  the submillimetre galaxies. Column 2 gives the ID number of the submillimetre
  galaxy as listed in Table~\ref{table:850cat} or Table~\ref{table:450cat}; we
  searched for counterparts to all 450-$\mu$m detected objects that were not
  detected at 850~$\mu$m.  Column 4 gives the catalogue number for each ID; for
  reference these numbers are also shown on Figs.~\ref{fig:post1} and
  \ref{fig:post2}. Column 8 gives the angular offset between the SCUBA position and
  the {\em Spitzer\/} position. Column 9 gives the corrected $P$ statistic (see
  text).}
\label{table:spit}
\vspace*{0.1in}
\begin{tabular}{lccccccccl}\hline
Field & ID & $\lambda$ & Cat\# & $S_{\lambda}$ & RA & Dec & $\Delta\theta$ & $P$ & Note\\ 
       & & ($\mu$m) & &($\mu$Jy) & (J2000.0) & (J2000.0)&  (arcsec)  &\\ \hline
RX\,J$005734.78-272827.4$ & $850\#1$ & $4.5$ & $272$ & $476\pm15$ & $00^{\rm h}$\
       $57^{\rm m}$\ $34^{\rm s}.93$  & $-27^{\circ}$\ $28^{\prime}$\
       $27^{\prime\prime}.6$ & $3.69$ & $0.005$ & QSO \\ [1pt]
& $850\#1$&  $8.0$ &  $203$ & $1534\pm31$ & $00^{\rm h}$\
       $57^{\rm m}$\ $34^{\rm s}.94$ & $-27^{\circ}$\ $28^{\prime}$\ 
       $27^{\prime\prime}.6$ & $3.67$ & $0.003$ & QSO \\ [1pt]
& $850\#1$ & $24.0$ & $10$ & $4107\pm111$ & $00^{\rm h}$\ $57^{\rm m}$\ $34^{\rm s}.90$
       & $-27^{\circ}$\ $28^{\prime}$\ $27^{\prime\prime}.3$ & $4.12$ & $0.002$
       & QSO \\ [1pt]
& $450\#2$ & $4.5$ & $231$ & $26.5\pm3.6$ & $00^{\rm h}$\ $57^{\rm m}$\ $33^{\rm s}.06$ &
       $-27^{\circ}$\ $28^{\prime}$\ $39^{\prime\prime}.7$ & $0.97$ & $0.009$
       & \\ [1pt]
& $450\#2$ & $8.0$ & $180$ & $18.4\pm4.4$ & $00^{\rm h}$\ $57^{\rm m}$\
       $33^{\rm s}.05$ & $-27^{\circ}$\ $28^{\prime}$\ $39^{\prime\prime}.6$ & $1.19$
       & $0.011$ \\ [5pt]

RX\,J$094144.51+385434.8$ & $850\#1$ & $4.5$ & $242$ & $713\pm19$ &  $09^{\rm h}$\
       $41^{\rm m}$\ $44^{\rm s}.69$ & $+38^{\circ}$\ $54^{\prime}$\
       $39^{\prime\prime}.8$ & $3.38$ & $0.003$ & QSO \\ [1pt]
& $850\#1$ & $8.0$ & $180$ & $2147\pm36$ & $09^{\rm h}$\ $41^{\rm m}$\ $44^{\rm s}.70$
       & $+38^{\circ}$\ $54^{\prime}$\ $39^{\prime\prime}.9$ & $3.41$ & $0.001$
       & QSO \\ [1pt]
& $850\#1$ & $24.0$ & $1$ & $5254\pm182$ & $09^{\rm h}$\ $41^{\rm m}$\ $44^{\rm s}.66$
       & $+38^{\circ}$\ $54^{\prime}$\ $39^{\prime\prime}.5$ & $3.30$ & $0.001$
       & QSO \\ [1pt]
& $850\#2$ & $4.5$ & $191$ & $87.9\pm6.6$ & $09^{\rm h}$\ $41^{\rm m}$\ $46^{\rm s}.34$ &
       $+38^{\circ}$\ $53^{\prime}$\ $56^{\prime\prime}.5$ & $2.80$ & $0.013$ &
       \\ [1pt]
& $850\#2$ & $8.0$ & $155$ & $65.7\pm6.9$ & $09^{\rm h}$\ $41^{\rm m}$\ $46^{\rm s}.33$ &
       $+38^{\circ}$\ $53^{\prime}$\ $55^{\prime\prime}.6$ & $2.45$ & $0.008$ &
       \\ [1pt]
& $850\#2$ & $24.0$ & $6$ &  $1041\pm166$ & $09^{\rm h}$\ $41^{\rm m}$\ $46^{\rm s}.38$
       & $+38^{\circ}$\ $53^{\prime}$\ $57^{\prime\prime}.0$ & $2.89$ & $0.004$
       & \\ [1pt]
& $850\#3$ & $4.5$ & $339$ & $74.7\pm6.0$ & $09^{\rm h}$\ $41^{\rm m}$\ $42^{\rm s}.66$ &
       $+38^{\circ}$\ $55^{\prime}$\ $36^{\prime\prime}.8$ & $2.53$ & $0.012$ &
       \\ [1pt]
& $850\#3$ & $8.0$ & $241$ & $63.9\pm6.9$ & $09^{\rm h}$\ $41^{\rm m}$\ $42^{\rm s}.68$ &
       $+38^{\circ}$\ $55^{\prime}$\ $36^{\prime\prime}.9$ & $2.50$ & $0.009$ &
       \\ [1pt]
& $450\#3$ & $4.5$ & $205$ & $19.0\pm 3.1$ &  $09^{\rm h}$\  $41^{\rm m}$\ $47^{\rm
       s}.00$ & $+38^{\circ}$\ $54^{\prime}$\ $11^{\prime\prime}.2$ & $4.90$ &
       $0.124$ & \\ [1pt] 
& $450\#3$ &  $4.5$ & $206$ & $13.1\pm2.6$ & $09^{\rm h}$\  $41^{\rm m}$\ $46^{\rm
       s}.39$ & $+38^{\circ}$\ $54^{\prime}$\ $13^{\prime\prime}.9$ & $4.11$ &
       $0.131$ & \\ [1pt]
& $450\#3$ & $8.0$ & $159$ & $24.9\pm4.8$ & $09^{\rm h}$\ $41^{\rm m}$\
       $46^{\rm s}.95$ & $+38^{\circ}$\ $54^{\prime}$\  $10^{\prime\prime}.9$ & $4.69$ 
       & $0.055$ \\ [1pt] 
& $450\#3$ & $8.0$ & $160$ & $22.3\pm4.7$ & $09^{\rm h}$\ $41^{\rm m}$\
       $46^{\rm s}.39$ & $+38^{\circ}$\ $54^{\prime}$\ $13^{\prime\prime}.9$ & $4.11$ 
       & $0.052$ \\ [5pt]

RX\,J$121803.82+470854.6$ & $850\#2$ & $4.5$ & $223$ & $47.8\pm4.9$ & $12^{\rm h}$\
       $18^{\rm m}$\ $07^{\rm s}.57$ & $+47^{\circ}$\ $08^{\prime}$\
       $45^{\prime\prime}.2$ & $0.92$ & $0.008$ & \\ [1pt]
& $850\#2$ & $8.0$ & $175$ & $35.9\pm5.3$ & $12^{\rm h}$\ $18^{\rm m}$\ $07^{\rm s}.57$ &
       $+47^{\circ}$\ $08^{\prime}$\ $45^{\prime\prime}.0$ & $0.90$  & $0.004$
       & \\ [1pt]
& $850\#2$ & $24.0$ & $6$ & $476\pm104$ & $12^{\rm h}$\ $18^{\rm m}$\ $07^{\rm s}.60$ &
       $+47^{\circ}$\ $08^{\prime}$\ $46^{\prime\prime}.8$ & $1.76$ & $0.001$ &
       \\ [1pt]
& $850\#4$ & $4.5$ & $256$ & $4.5\pm1.5$ & $12^{\rm h}$\ $18^{\rm m}$\ $02^{\rm s}.54$ &
       $+47^{\circ}$\ $08^{\prime}$\ $28^{\prime\prime}.3$ & $4.07$ & $0.317$ &
       \\ [1pt] 
& $850\#4$ & $8.0$ & $188$ & $16.7\pm4.0$ & $12^{\rm h}$\ $18^{\rm m}$\
       $02^{\rm s}.67$ & $+47^{\circ}$\ $08^{\prime}$\ $27^{\prime\prime}.2$ &
       $2.48$ & $0.038$ \\ [1pt]
& $850\#5$ & $ 24.0$ & $5$ & $3200\pm160$ & $12^{\rm h}$\ $18^{\rm m}$\ $04^{\rm s}.46$
       & $+47^{\circ}$\ $08^{\prime}$\ $50^{\prime\prime}.6$ & $8.87$ & $0.006$
       & likely false ID\\ [5pt]

RX\,J$124913.86-055906.2$ & $850\#1$ & $4.5$ & $261$ & $3671\pm42$ & $12^{\rm
  h}$\ $49^{\rm m}$\ $13^{\rm s}.88$ & $-05^{\circ}$\ $59^{\prime}$\
  $19^{\prime\prime}.0$ & $1.71$ & $<0.001$ & QSO \\ [1pt] 
& $850\#1$ & $8.0$ & $200$ & $12764\pm88$ & $12^{\rm h}$\ $49^{\rm m}$\
  $13^{\rm s}.90$ & $-05^{\circ}$\ $59^{\prime}$\ $18^{\prime\prime}.9$ &
  $1.46$ & $<0.001$ & QSO\\ [1pt]
& $850\#1$ & $24.0$ & $5$ & $26311\pm125$ & $12^{\rm h}$\ $49^{\rm m}$\ $13^{\rm
  s}.86$ & $-05^{\circ}$\ $59^{\prime}$\ $18^{\prime\prime}.4$ & $2.13$ &
  $0.001$ & QSO \\ [1pt] 
& $850\#4$ & $4.5$ & $342$ & $18.1\pm3.0$ & $12^{\rm h}$\ $49^{\rm m}$\
  $12^{\rm s}.56$ & $-05^{\circ}$\ $58^{\prime}$\ $34^{\prime\prime}.9$ &
  $3.18$ & $0.093$ & \\ [1pt]
& $850\#4$ & $4.5$ & $343$ & $79.1\pm6.2$ & $12^{\rm h}$\ $49^{\rm m}$\
  $12^{\rm s}.26$ & $-05^{\circ}$\ $58^{\prime}$\ $30^{\prime\prime}.6$ &
  $3.34$ &  $0.018$ &\\ [1pt]
& $850\#4$ & $8.0$ & $265$ & $55.9\pm7.5$ & $12^{\rm h}$\ $49^{\rm m}$\
  $12^{\rm s}.26$ & $-05^{\circ}$\ $58^{\prime}$\ $31^{\prime\prime}.2$ &
  $2.82$ & $0.015$ &\\ [1pt]
& $850\#5$ & $4.5$ & $324$ & $25.5\pm3.6$ & $12^{\rm h}$\ $49^{\rm m}$\
  $11^{\rm s}.39$ & $-05^{\circ}$\ $59^{\prime}$\ $57^{\prime\prime}.9$ &
  $2.89$ & $0.061$ &\\ [1pt]
& $850\#5$ & $8.0$ & $236$ & $33.7\pm6.6$ & $12^{\rm h}$\ $49^{\rm m}$\
  $11^{\rm s}.44$ & $-05^{\circ}$\ $59^{\prime}$\ $58^{\prime\prime}.4$ &
  $2.00$ & $0.023$ &\\ [1pt]
& $450\#2$ & $4.5$ & $404$ & $2.2\pm1.1$ & $12^{\rm h}$\ $49^{\rm m}$\
  $10^{\rm s}.95$ & $-05^{\circ}$\ $58^{\prime}$\ $34^{\prime\prime}.7$ &
  $3.06$ & $0.194$ &\\ [5pt] 

RX\,J$163308.57+570258.7$ & $850\#1$ & $4.5$ & $142$ & $21.3\pm3.3$ & $16^{\rm h}$\
       $32^{\rm m}$\ $59^{\rm s}.74$ & $+57^{\circ}$\ $03^{\prime}$\
       $21^{\prime\prime}.5$ & $3.48$ & $0.076$ & \\ [1pt] 
& $850\#1$ & $8.0$ & $134$ & $30.0\pm5.3$ & $16^{\rm h}$\ $32^{\rm m}$\ $59^{\rm s}.74$ &
       $+57^{\circ}$\ $03^{\prime}$\ $21^{\prime\prime}.3$ & $3.29$ & $0.037$ &
       \\ [1pt]
& $850\#1$ & $24.0$ & $17$ & $275\pm55$ & $16^{\rm h}$\ $32^{\rm m}$\ $59^{\rm s}.65$
       & $+57^{\circ}$\ $03^{\prime}$\ $20^{\prime\prime}.8$ & $3.47$ & $0.017$
       & \\ [1pt]
& $850\#2$ & $4.5$ & $242$ & $83.3\pm6.4$ & $16^{\rm h}$\ $33^{\rm m}$\ $08^{\rm s}.55$ &
       $+57^{\circ}$\ $02^{\prime}$\ $54^{\prime\prime}.6$ & $4.67$ & $0.038$ &
       QSO \\ [1pt]
& $850\#2$ & $8.0$ & $226$ & $153\pm10$ & $16^{\rm h}$\ $33^{\rm m}$\ $08^{\rm s}.52$ &
       $+57^{\circ}$\ $02^{\prime}$\ $54^{\prime\prime}.5$ & $4.86$ & $0.023$ &
       QSO \\ [1pt]
& $850\#2$ & $24.0$ & $18$ & $705\pm40$ & $16^{\rm h}$\ $33^{\rm m}$\ $08^{\rm s}.54$
       & $+57^{\circ}$\ $02^{\prime}$\ $53^{\prime\prime}.0$ & $6.18$ & $0.004$
       & QSO \\ [1pt]
& $850\#5$ & $4.5$ & $174$ & $19.1\pm3.1$ & $16^{\rm h}$\ $33^{\rm m}$\ $03^{\rm s}.17$ &
       $+57^{\circ}$\ $02^{\prime}$\ $58.^{\prime\prime}6$ & $0.13$ & $<0.001$ &
       \\ [1pt]
& $850\#5$ & $8.0$ & $168$ & $17.0\pm4.5$ & $16^{\rm h}$\ $33^{\rm m}$\ $03^{\rm s}.18$ &
       $+57^{\circ}$\ $02^{\prime}$\ $58^{\prime\prime}.9$ & $0.16$ & $<0.001$ &
       \\ [1pt]
& $850\#5$ & $24.0$ & $19$ & $256\pm32$ & $16^{\rm h}$\ $33^{\rm m}$\ $03^{\rm s}.09$
       & $+57^{\circ}$\ $02^{\prime}$\ $57^{\prime\prime}.6$ & $1.27$ & $0.002$
       & \\ [1pt]
& $850\#6$  & $4.5$ & $290$ & $2.9\pm1.3$ & $16^{\rm h}$\ $33^{\rm m}$\ $12^{\rm s}.45$ &
       $+57^{\circ}$\ $02^{\prime}$\ $34^{\prime\prime}.9$ & $3.20$ & $0.172$ & \\ [1pt]
& $850\#7$  & $4.5$ & $225$ & $25.3\pm3.6$ & $16^{\rm h}$\ $33^{\rm m}$\
       $05^{\rm s}.23$ & $+57^{\circ}$\ $04^{\prime}$\ $17^{\prime\prime}.1$ &
       \ldots & \ldots & See Appendix A \\ [1pt]
& $850\#7$  & $4.5$ & $243$ & $17.1\pm2.9$ & $16^{\rm h}$\ $33^{\rm m}$\
       $06^{\rm s}.23$ & $+57^{\circ}$\ $04^{\prime}$\ $09^{\prime\prime}.5$ &
       \ldots & \ldots & See Appendix A \\ [1pt]
& $850\#7$  & $4.5$ & $260$ & $29.8\pm3.9$ & $16^{\rm h}$\ $33^{\rm m}$\
       $06^{\rm s}.30$ & $+57^{\circ}$\ $04^{\prime}$\ $03^{\prime\prime}.2$ &
       \ldots & \ldots & See Appendix A \\ [1pt]
& $850\#7$  & $8.0$ & $232$ & $58.7\pm6.7$ & $16^{\rm h}$\ $33^{\rm m}$\
       $06^{\rm s}.29$ & $+57^{\circ}$\ $04^{\prime}$\ $03^{\prime\prime}.3$ &
       \ldots & \ldots & See Appendix A \\ [1pt]  
& $850\#7$  & $8.0$ & $229$ & $21.5\pm4.8$ & $16^{\rm h}$\ $33^{\rm m}$\
       $06^{\rm s}.24$ & $+57^{\circ}$\ $04^{\prime}$\ $09^{\prime\prime}.9$ &
       \ldots & \ldots & See Appendix A \\ [1pt]  
& $850\#7$  & $8.0$ & $212$ & $20.4\pm4.7$ & $16^{\rm h}$\ $33^{\rm m}$\
       $05^{\rm s}.34$ & $+57^{\circ}$\ $04^{\prime}$\ $17^{\prime\prime}.6$ &
       \ldots & \ldots & See Appendix A \\ [1pt]  
& $850\#7$  & $8.0$ & $211$ & $18.6\pm4.6$ & $16^{\rm h}$\ $33^{\rm m}$\
       $04^{\rm s}.96$ & $+57^{\circ}$\ $04^{\prime}$\ $15^{\prime\prime}.7$ &
       \ldots & \ldots & See Appendix A \\ [1pt]  
& $850\#7$  & $8.0$ & $216$ & $13.6\pm4.3$ & $16^{\rm h}$\ $33^{\rm m}$\
       $05^{\rm s}.52$ & $+57^{\circ}$\ $04^{\prime}$\ $11^{\prime\prime}.8$ &
       \ldots & \ldots & See Appendix A \\ [1pt]  
& $850\#7$  & $24.0$ & $8$ & $249\pm49$ & $16^{\rm h}$\ $33^{\rm m}$\ $05^{\rm
       s}.05$ & $+57^{\circ}$\ $04^{\prime}$\ $14^{\prime\prime}.5$ & \ldots &
       \ldots & See Appendix A \\ [1pt]  
& $850\#7$  & $24.0$ & $10$ & $270\pm49$ & $16^{\rm h}$\ $33^{\rm m}$\ $06^{\rm
       s}.28$ & $+57^{\circ}$\ $04^{\prime}$\ $02^{\prime\prime}.6$ & \ldots &
       \ldots & See Appendix A \\ 
\hline 
\end{tabular}
\end{table*}

\section{Mid-infrared colours}

\label{section:midirc}

\begin{figure*}
\setlength{\unitlength}{1in}
\begin{picture}(8.0,4.3)
\includegraphics{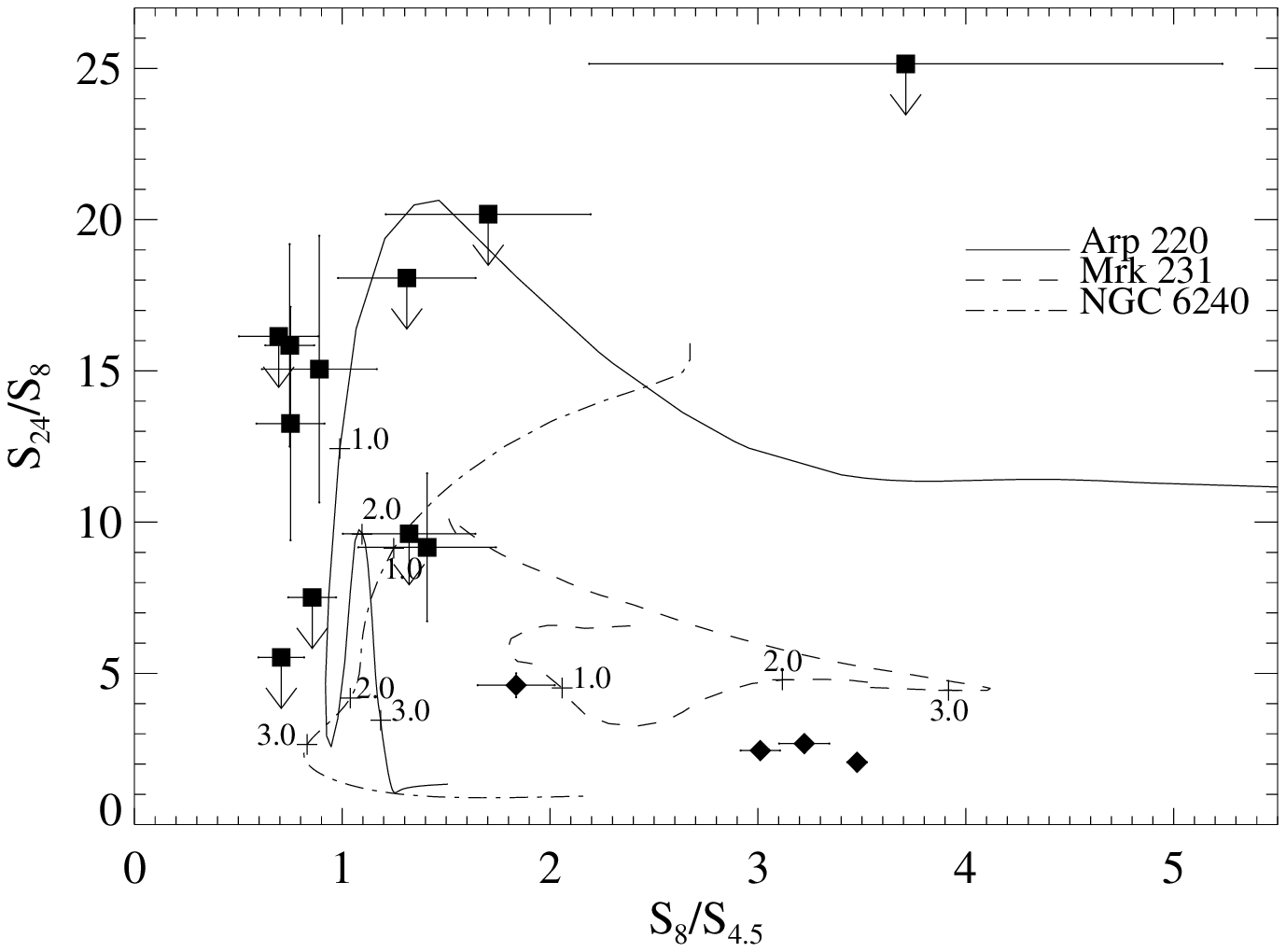}
\end{picture}
\caption[dum]{Colour-colour plot for those sources detected in two or three of
the observed {\em Spitzer\/} bands (24, 8.0 and $4.5~\mu$m). The QSOs are
plotted as diamonds while the submillimetre-selected galaxies discovered in
their fields are shown as squares. Lines show the redshifted colour tracks of
three local ultraluminous infrared galaxies with (Mrk 231) and without (Arp
220 and NGC 6240) obvious signatures of powerful AGNs at mid-infrared
wavelengths. Redshifts of 1.0, 2.0 and 3.0 are marked on each of these
tracks. }
\label{fig:colours}
\end{figure*}

Mid-infrared colours provide a powerful means of distinguishing between galaxies which contain AGN and those that do not (e.g. Ivison et al. 2004, 2007; Egami et al. 2004). Luminous galaxies at $1<z<3$ hosting AGN have power-law mid-infrared spectra whereas those without AGN are dominated by starlight in the IRAC bands and dust/PAH emission at 24~$\mu$m.

In Fig.~\ref{fig:colours} we plot the ratio of 24 and 8~$\mu$m flux density against the ratio of 8 and 4.5~$\mu$m flux density for those objects detected in at least two {\em Spitzer\/} bands. We superimpose redshift tracks for the local ULIRGs Mrk 231 (AGN-like SED) and Arp~220 and NGC~6240 (starburst-like SEDs). The QSOs and SMGs clearly inhabit a different regions of this diagram. As expected the QSOs have colours more similar to Mrk~231 than to the starburst galaxies. Three QSOs have remarkably similar mid-infrared colours given that they were selected on their X-ray properties. The other QSO, RX~J163308, has significantly different IRAC colours but we note from the postage stamp image (Fig.~\ref{fig:post1}) that it has a close `companion' galaxy which is blended with the QSO in our extracted catalogue and will thus contaminate the colours. While the other three QSOs have $8.0/4.5~\mu$m flux density ratios roughly consistent with the Mrk~231 template at their known redshifts we note that their 24/8.0~$\mu$m flux density ratios appear to be lower than that of Mrk~231 by a factor of $\sim2$ which might be related to the relative strength of emission/absorption features. All but one of the SMGs on the other hand have mid-infrared colours consistent with Arp 220 and/or NGC 6240 at $1<z<3$. The one exception has colours consistent with Arp~220 at $0.1<z<0.5$ but its faintness and non-detection at 24~$\mu$m are more consistent with a Mrk 231-like object at $1<z<3$. It is thus the only SMG in our sample for which mid-infrared colours indicate significant emission from an AGN. We thus find no strong evidence for buried AGN in the SMGs.  Having said this, we note that deep X-ray (e.g. Vignati et al. 1999; Netzer et al. 2005) and infrared spectroscopy (e.g. Lutz et al. 2003; Armus et al. 2006) observations of NGC~6240 show that this galaxy does in fact host a highly obscured AGN but that the energetics are dominated by the starburst. This leaves open the possibility that many of the SMGs contain buried AGN which are not powerful enough to dominate over the starburst in the mid-infrared.

\section{Discussion and conclusions}

We have shown that the number counts of SMGs with flux densities $\sim2-4$~mJy in the $\sim 2$~arcmin diameter fields surrounding our X-ray absorbed QSOs is higher than that found for a blank field. Most of these galaxies have mid-infrared colours very similar to the local ULIRG Arp~220 if it were placed at $1<z<3$. Fitting the millimetre to far-infrared spectrum of Arp~220 with a greybody and then integrating to determine the far-infrared luminosity we find L$_{\rm FIR}=(1.3\pm0.1) \times 10^{12}$~L$_{\odot}$ which corresponds to a star-formation rate (SFR) of $224\pm17$~M$_{\odot}$~yr$^{-1}$ after applying the usual scaling-law (Kennicutt 1998). Using the same greybody fit and applying the appropriate k-correction leads to predicted 850~$\mu$m flux densities of 1.73, 1.44 and 1.42 mJy for Arp~220 at z=1, 2 and 3 respectively. The SMGs in our sample have 850~$\mu$m flux densities ranging from about $3-10$ mJy implying that their SFRs range between $400-1300$~M$_{\odot}$~yr$^{-1}$ if Arp~220 is indeed a good analogue. Therefore, if the starburst can be sustained for a few hundred million years or so then a stellar mass equivalent to that of the bulge of a large galaxy will be assembled.

It is natural to speculate that an over-dense region of space at high redshift containing galaxies that are capable of building large, homogeneous populations of stars will evolve over cosmic time to become clusters of galaxies containing a population of massive elliptical galaxies. Since it is well established that elliptical galaxies contain dormant SMBHs (e.g. Magorrian et al. 1998), and that these must have assembled at high redshifts where the accretion luminosity density peaks, we expect that our SMGs should contain growing black holes, i.e. AGN. How the major growth phase of the black hole coincides temporally with the luminous starburst that builds up the stellar mass is a current topic of debate. Indeed, models predict a range of possibilities (e.g. Archibald et al 2002; Di Matteo, Springel \& Hernquist 2005) but there must be a link between the formation of the stars in the bulge and the SMBH because of the tight correlation between their masses found for local galaxies (Ferrarese \& Merritt 2000). After a number of early null results using moderate depth X-ray data (e.g. Fabian et al. 2000; Waskett et al. 2003) deep X-ray (Alexander et al. 2005a,b, 2008) and mid-infrared spectroscopy (Men\'{e}ndez-Delmestre et al. 2007; Pope et al. 2008) observations show that SMGs found in the field often do contain AGN but that the masses of their black holes may be much smaller than those found in high redshift QSOs, and consequently that the bulk of the far-infrared emission is due to the starburst. Moreover, estimates of the host galaxy masses suggest that the growth of the black hole lags the growth of the spheroid (Borys et al. 2005; Alexander et al. 2008; see also Coppin et al. 2009). While this may be the case for SMGs, analysis of massive galaxies selected on other criteria find evidence to the contrary, suggesting that more than one scenario is possible for the formation of a massive galaxy and its SMBH. For example, studies of powerful radio sources (McLure et al. 2006) and luminous QSOs (Peng et al. 2006) suggest that the black-hole mass to bulge luminosity ratio increases towards high redshifts thus indicating that the black hole reaches its critical mass before the galaxy bulge is fully formed.

The mid-infrared colours of our SMGs also argue against the presence of a powerful AGN in these objects unless they are very highly obscured. They are thus either intrinsically different objects to the QSO or they are caught at a different time in their evolution. Page et al. (2010, in prep.) argue that the X-ray absorption in the QSOs has its origin in an ionized absorber associated with a powerful wind capable of terminating the star formation. Such feedback is a popular ingredient of models since it leads naturally to the relationship between SMBH mass and spheroid mass discussed above (e.g. Fabian 1999). If the SMGs in the same field evolve to form a cluster of ellipticals then they would presumably go through a similar evolutionary sequence to the QSO but are caught at an earlier stage of their life cycle at $1<z<3$. Estimates of the stellar masses would be useful to put further constraints on their evolutionary status. To this end we are currently fitting models to optical through infrared data to determine photometric redshifts and restframe $K$-band luminosities; an analysis of the RXJ~094144 field is currently in progress (Carrera et al. 2010, in prep.).

One difference worthy of note between the findings of this paper and those for high-redshift (typically $z>3$) radio galaxies (HzRGs) concerns very bright SMGs in their fields. Two of the seven HzRG fields imaged with SCUBA contain SMGs with $S_{850}>20$ mJy (Stevens et al. 2003) whereas the brightest object found around the QSOs has $S_{850}\sim10$~mJy. Although the samples are too small to give a statistically significant result it appears that the SMGs detected around the lower redshift, less powerful, radio-quiet AGN have lower $S_{850}$. A direct comparison is not informative because the sources were extracted with different signal-to-noise criteria and by different methods but comparing flux densities significantly above the catalogue cut-offs yields $8$ SMGs with $S_{850}\geq6$~mJy in the 7 HzRG fields while only $2$ are detected around the 5 QSOs.

Given the effectively flat relationship between $S_{850}$ and redshift for $1<z<10$ this difference in flux density leads to a proportional difference in star-formation rates (all other things being equal). There is thus reasonably good evidence for a difference in star-formation activity in individual galaxies in these fields indicating that the galaxy formation process might be dependent on the magnitude of the density peak and/or its redshift. For example, in the case of HzRGs it was found that the detection rate increased dramatically at $z>2.5$ (Archibald et al. 2001; Reuland et al. 2004) which is significantly higher than the average redshift of our QSO sample.  Another possibility is that the radio properties of the central AGN are in some way linked to environmental density (e.g. Kauffmann, Heckman \& Best 2008). Indeed, recent work by Falder et al. (2010) finds evidence for larger over-densities of galaxies around radio-loud objects (based on work carried out at 3.6~$\mu$m). Perhaps radio synchrotron luminosity is enhanced in regions where the inter-galactic medium is more dense.

An obvious next step will be to conduct a statistically meaningful survey of fields around AGN over the full range of redshifts in order to track star-formation activity as a function of cosmic time. The next generation bolometer camera on the JCMT (SCUBA-2) and SPIRE/PACS on-board the {\em Herschel Space Observatory\/} are instruments with good enough sensitivity to make such a campaign practical. It would also allow a bigger area to be mapped around each AGN, better matching the size of proto-cluster regions predicted by numerical simulations.

\section*{ACKNOWLEDGMENTS}

We thank Ian Smail for extensive comments on the draft manuscript and Mark Thompson for useful discussions.  The James Clerk Maxwell Telescope is operated by The Joint Astronomy Centre on behalf of the Science and Technology Facilities Council of the United Kingdom, the Netherlands Organisation for Scientific Research, and the National Research Council of Canada. JCMT data were taken under project IDs M03AU46, M03BU32 and M04BU14.  This work is based [in part] on observations made with the Spitzer Space Telescope, which is operated by the Jet Propulsion Laboratory, California Institute of Technology under a contract with NASA. Support for this work was provided by NASA through an award issued by JPL/Caltech. J.A.S., M.J.P. and F.J.C. acknowledge support from the Royal Society. F.J.C. acknowledges further support from the Spanish Ministerio de Educaci\'on y Ciencia under project ESP2006-13608.

\appendix

\section[]{Notes and postage stamp images}

In this section we present and discuss postage stamp images of the QSOs and SMGs. For the RX~J$094144$ field we compare the mid-infrared counterparts to the SMGs with those identified previously at optical and near-infrared wavelengths. We also discuss cases where likely mid-infrared counterparts were not catalogued because they fell just outside our formal search radii, and one case where an obvious SMG was not found by our extraction algorithm; these sources were not included in the analysis described in the main part of this paper.

\subsection{The RX~J094144.51+385434.8 field}

The SCUBA data for this field were published previously by Stevens et al. (2004). In the present paper we have treated the data slightly differently in the final reduction stages (rebinning and smoothing). We also apply a formal source extraction algorithm in this work resulting in only 2 sources at 850~$\mu$m and 4 at 450~$\mu$m whereas Stevens et al. (2004) simply identified peaks on the maps above a 3-$\sigma$ signal-to-noise threshold, identifying 6 sources. As already noted in Section~\ref{section:se} the source extraction algorithm fails to detect sources that are blended which has undoubtedly affected the 850~$\mu$m extraction. At 450~$\mu$m we can see that sources 450\#1 and 450\#3 appear more elongated than the other two objects hinting that they may be resolved. It was these objects that Stevens et al. (2004) identified as blends.

Inspection of the postage stamps shows that source {\bf 850\#2/450\#2} is coincident with a bright mid-infrared source detected in all 3 {\em Spitzer\/} channels. This object was one of a pair of sources detected in the $R$- and $K$-bands by Stevens et al. (2004). The new {\em Spitzer\/} data show that the correct identification is the object with the redder $R-K$ colour since this object becomes progressively more prominent with increasing wavelength compared to its neighbour and is detected at 24~$\mu$m.

Source {\bf 450\#3} falls in a crowded field. We find 2 possible identifications in the IRAC channels. Of these source 206/160 lies closer to the 450~$\mu$m source centroid and is also coincident with faint (formally undetected) emission at 24~$\mu$m. Source 206/160 was not one of the possible identifications discussed by Stevens et al. (2004) since it is very faint/undetected in the $K$-band. We conclude that the likely identification for this SMG is the very red object 206/160.

Source {\bf 450\#4} was matched with an extremely red object ($R-K=5.8$) by Stevens et al. (2004). This source is not formally recovered in this work because it lies outside our 5 arcsec IRAC search radius but is visible in the postage stamp to the west of the 450~$\mu$m centroid as a bright source at 4.5 and 8.0~$\mu$m. Given its red colours we consider this object as a likely identification. Measured flux densities are $57.0\pm5.3$ and $48.5\pm6.1$ $\mu$Jy at 4.5 and 8.0~$\mu$m respectively giving a colour similar to Arp~220 at the redshift of the QSO. The 4.5-$\mu$m centroid is 6.6 arcsec from the extracted 450~$\mu$m position at $09^{\rm h}\ 41^{\rm m}\ 43^{\rm s}.61$, $+38^{\circ}\ 54^{\prime}\ 24^{\prime\prime}.5$ while the 8.0-$\mu$m centroid is 6.4 arcsec from the extracted 450~$\mu$m position at $09^{\rm h}\ 41^{\rm m}\ 43^{\rm s}.64$, $+38^{\circ}\ 54^{\prime}\ 24^{\prime\prime}.6$. The calculated P-statistic is 0.089 and 0.064 at 4.5 and 8.0~$\mu$m respectively.

\subsection{The RX~J121803.82+470854.6 field}

In Section~\ref{section:catmem} we noted that the QSO in this field was not found by our extraction algorithm but was detected by Page et al. (2001). The latter observations used SCUBA in `photometry' mode where the central bolometer of each array is centred on the target but no image is made.  Inspection of Fig.~\ref{fig:allmaps} shows that while there is submillimetre emission at the location of the QSO (located below source 5 on the 850~$\mu$m image) the surrounding structure is complex. It is thus possible that the source extraction algorithm failed to locate a significant peak because of blending. It is also likely that the photometry observation of this object reported by Page et al. (2001) is contaminated by emission from the SMGs close to the QSO.

In Table~\ref{table:spit} we noted that object {\bf 850\#5} has a likely false identification at 24~$\mu$m. This can also be attributed to the crowded field close to the QSO since it is the QSO itself that has fallen within the search radius at 24~$\mu$m while it was not picked up in the smaller search radii adopted for the IRAC channels.

\subsection{The RX~J124913.86$-$055906.2 field}

Source {\bf 850\#2} has no listed identification in Table~\ref{table:spit} but
the postage stamp images show bright 4.5 and 8.0~$\mu$m sources close to the
850~$\mu$m centroid. The closest 4.5~$\mu$m source is at $12^{\rm h}\ 49^{\rm
  m}\ 15^{\rm s}.19$, $-05^{\circ}\ 58^{\prime}\ 40^{\prime\prime}.9$ which at 5.3 arcsec from the SMG is only just outside our formal search radius. This object has a flux density of $17.7\pm3.0$~$\mu$Jy and is coincident with an 8.0~$\mu$m source although the latter is not detected by {\sc SExtractor}. Inspection of the 8.0~$\mu$m image reveals that the diffraction spike from the bright QSO at the field centre falls on top of this object probably explaining the non-detection. Although the P-statistic for the 4.5~$\mu$m source is high at 0.259 we consider this object as a likely identification because of its detection in both IRAC bands.

Sources {\bf 850\#4} and {\bf 850\#4} have an obvious identification in the IRAC bands as well as faint (formally undetected) emission at 24~$\mu$m.

\subsection{The RX~J163308.57+570258.7 field}

Source {\bf 850\#4} has no listed identification in Table~\ref{table:spit} but the postage stamp images show a bright 4.5 and 8.0~$\mu$m source close to the 850~$\mu$m centroid along with faint (formally undetected) emission at 24~$\mu$m. The 4.5~$\mu$m source is located at $16^{\rm h}\ 33^{\rm m}\ 10^{\rm s}.15$, $+57^{\circ}\ 02^{\prime}\ 10^{\prime\prime}.2$ only 5.2 arcsec from the SMG, has a flux density of $12.9\pm2.6$ and a P-statistic of 0.266. The 8.0~$\mu$m source is located at $16^{\rm h}\ 33^{\rm m}\ 10^{\rm s}.08$, $+57^{\circ}\ 02^{\prime}\ 09^{\prime\prime}.3$ only 5.9 arcsec from the SMG, has a flux density of $17.6\pm4.5$ and a P-statistic of 0.198. This object has IRAC colours similar to Arp~220 at the redshift of the QSO.  The proximity of this object to the SMG together with its detection in in both IRAC bands make it a likely identification.

Inspection of the 850~$\mu$m map for this field (Fig.~\ref{fig:allmaps}) shows submillimetre emission north of the QSO at approximately $16^{\rm h}\ 33^{\rm m}\ 05^{\rm s}.75$, $+57^{\circ}\ 04^{\prime}\ 11^{\prime\prime}.7$. There is a negative region to the east of this position which is consistent with a negative image of the source located at one of the off-positions. The western off-position would fall off the map. We believe that this is a real SMG (probably two sources) which was missed by the extraction algorithm because of its proximity to the edge of the map. Inspection of the postage stamp images reinforces this belief since we find two 24~$\mu$m sources coincident with the SMG position and at the same approximate position angle as the 850~$\mu$m emission. The IRAC data reveal 3 and 5 sources at 4.5 and 8.0~$\mu$m respectively. The southern 24/850~$\mu$m source has two possible IRAC identifications. Of these, source 260/232 is the brightest and reddest object; its $8.0/4.5~\mu$m flux density ratio is $1.96\pm0.22$. If source 10 is the 24~$\mu$m counterpart then the ratio of flux densities at 24 and 8.0~$\mu$m is $4.60\pm0.99$ which would place this object on the Mrk~231 curve shown in Fig~\ref{fig:colours} indicating that it hosts a buried AGN, possibly at a lower redshift than RX~J163308.  The northern 24/850~$\mu$m source has 1 counterpart at 4.5~$\mu$m and 3 at 8.0~$\mu$m although the 4.5~$\mu$m emission is extended in the same manner as that at 8.0~$\mu$m implying that this SMG could have multiple mid-infrared counterparts. Inspection of the flux densities presented in Table~\ref{table:spit} shows that the IRAC colours are similar to those of Arp~220 at the redshift of the QSO so this source is unlikely to host an AGN.  Since this system is one of the most interesting in our sample we include it in Table~\ref{table:spit} as source {\bf 850\#7} although it is not included in the 850~$\mu$m number counts analysis because it was not found by the extraction algorithm.

\begin{figure*}
\setlength{\unitlength}{1in}
\begin{picture}(3.5,8.7)
\includegraphics{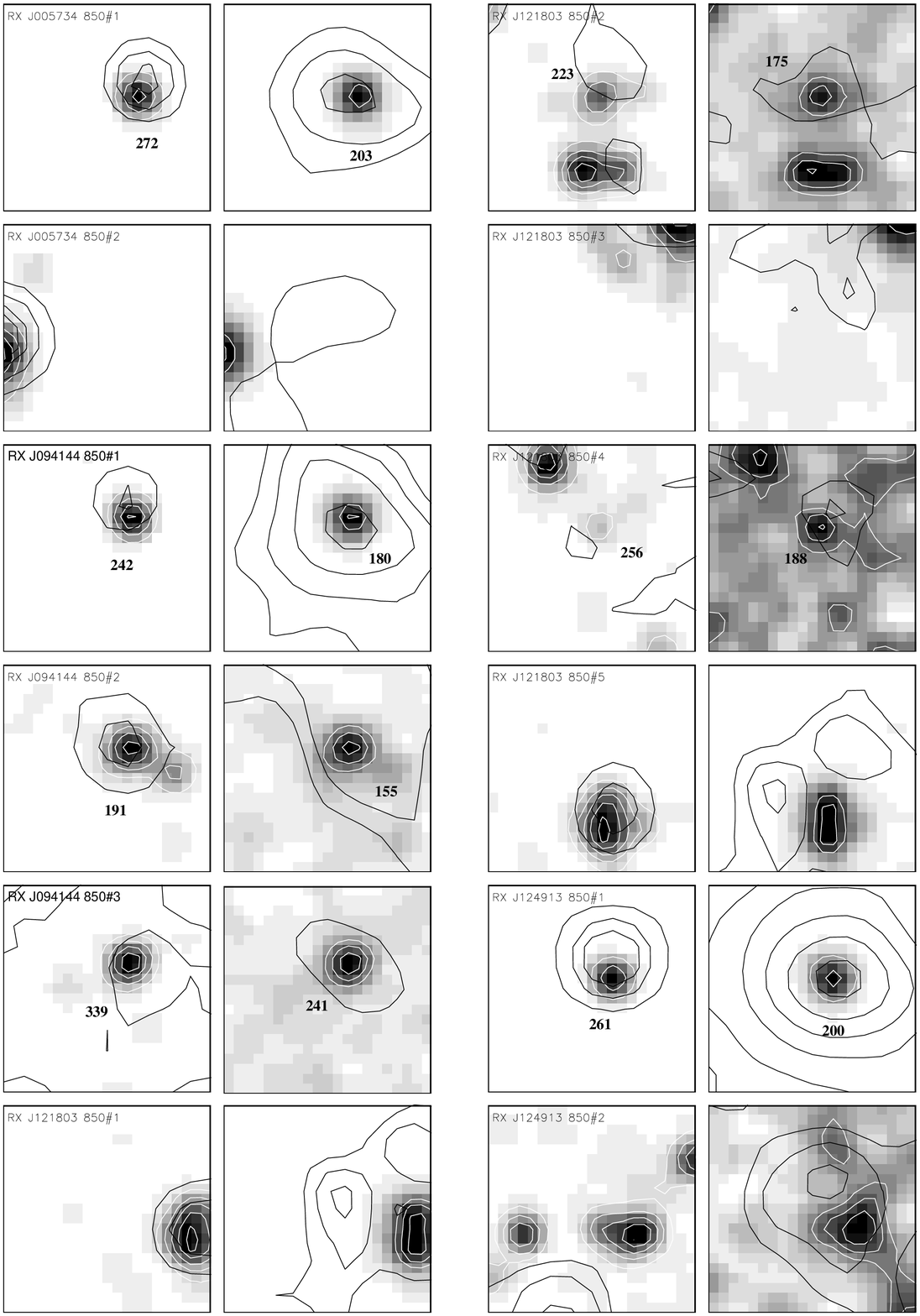}
\end{picture}
\caption[dum]{Postage stamps of $25\times25$ arcsec cut out at the extracted
  right ascension and declination of the 850~$\mu$m sources from Table\
  \ref{table:850cat}. The panels show (left) MIPS 24~$\mu$m contours (black)
  overlaid on the IRAC 4.5~$\mu$m data (grey-scale and white contours) and
  (right) SCUBA 850~$\mu$m contours (black) overlaid on the IRAC 8.0~$\mu$m
  data (grey-scale and white contours). Contour levels at 850~$\mu$m start at
  $\sim2~\sigma$ and increase in $\sim1~\sigma$ steps while the {\em Spitzer\/}
  contours are chosen to highlight faint sources that might not have been
  formally extracted.  The numbering corresponds to the catalogue numbers given
  in Table~\ref{table:spit}.}
\label{fig:post1}
\end{figure*}

\addtocounter{figure}{-1}

\begin{figure*}
\setlength{\unitlength}{1in}
\begin{picture}(3.5,7.5)
\includegraphics{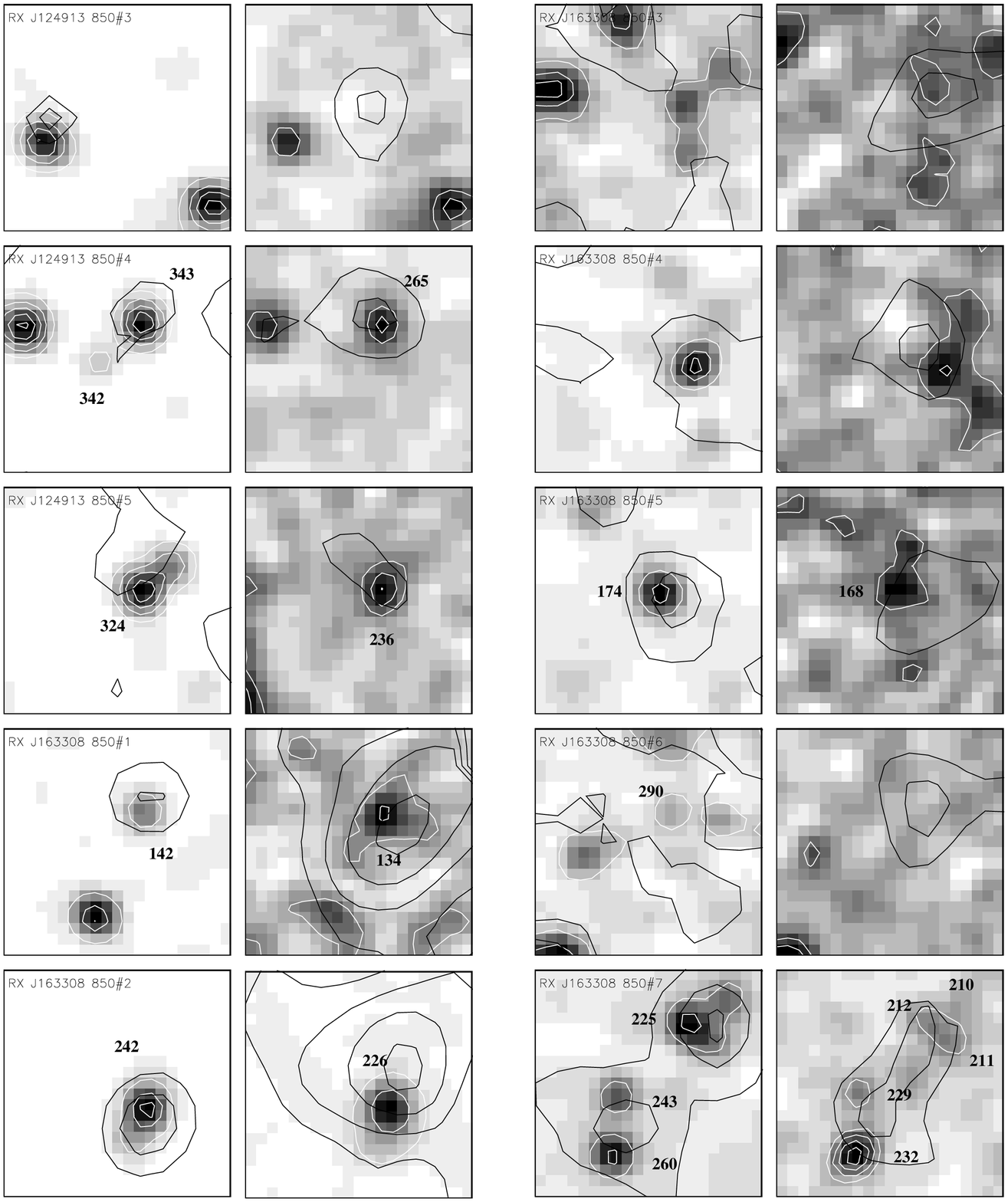}
\end{picture}
\caption[dum]{- continued}
\label{fig:post1}
\end{figure*}

\begin{figure*}
\setlength{\unitlength}{1in}
\begin{picture}(3.5,4.7)
\includegraphics{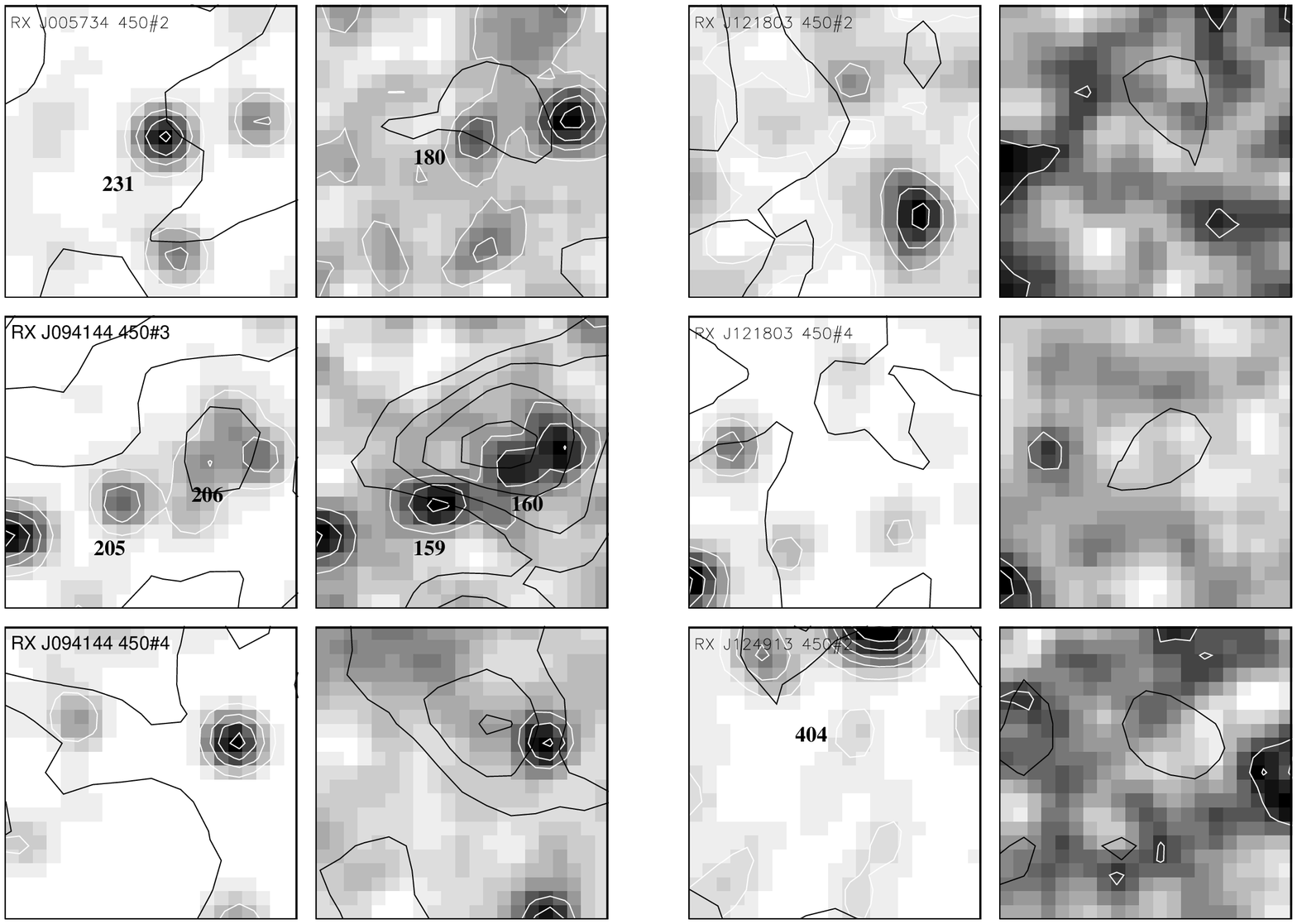}
\end{picture}
\caption[dum]{Postage stamps of $25\times25$ arcsec cut out at the extracted
  right ascension and declination of the 450~$\mu$m sources from Table\
  \ref{table:450cat}. The panels show (left) MIPS 24~$\mu$m contours (black)
  overlaid on the IRAC 4.5~$\mu$m data (grey-scale and white contours) and
  (right) SCUBA 450~$\mu$m contours (black) overlaid on the IRAC 8.0~$\mu$m
  data (grey-scale and white contours). Contour levels at 450~$\mu$m start at
  $\sim2~\sigma$ and increase in $\sim1~\sigma$ steps while the {\em Spitzer\/}
  contours are chosen to highlight faint sources that might not have been
  formally extracted.  The numbering corresponds to the catalogue numbers given
  in Table~\ref{table:spit}.}
\label{fig:post2}
\end{figure*}

\bsp

\end{document}